\documentclass[aps,pra,10pt,twocolumn,groupedaddress,notitlepage,showpacs,floatfix,longbibliography]{revtex4-1}
\usepackage{graphicx,times,bm,bbm,amssymb,amsmath,amsfonts,amsthm,mathrsfs,MnSymbol,float,tikz,physics,xcolor}
\usetikzlibrary{shapes.geometric, arrows}
\tikzstyle{arrow} = [thick,->,>=stealth]
\usepackage{svg}
\usepackage{booktabs}
\usepackage[pdfstartview=FitH]{hyperref}
\usepackage{comment}

\hypersetup{
    colorlinks=true,
    linkcolor=red,
    citecolor=magenta,
    filecolor=magenta,
    urlcolor=cyan,
    runcolor=cyan
}

\tikzstyle{startstop} = [rectangle, rounded corners, 
minimum width=1.5cm, 
minimum height=1cm,
text centered, 
draw=black, 
fill=none]
\tikzstyle{process} = [rectangle, rounded corners, 
minimum width=2cm, 
minimum height=1cm,
text centered, 
draw=black, 
fill=none]
\tikzstyle{io} = [ellipse, 
trapezium stretches=true,
trapezium left angle=70, 
trapezium right angle=110, 
minimum width=2cm, 
minimum height=1cm, text centered, 
draw=black, fill=none]

\begin{document}

\title{End-to-end entanglement of quantum network paths with multi-parameter states}
\author{Md Sohel Mondal, Aniket Zambare, Siddhartha Santra}
\affiliation{Department of Physics and Center of excellence in Quantum information, computation, science and technology, Indian Institute of Technology Bombay, Mumbai, Maharashtra 400076, India}
\begin{abstract}
Long-range entanglement distribution in a quantum network relies on entanglement swapping at intermediate nodes along a network path to connect short-range entangled states established over the network edges. The end-to-end entanglement of a network path obtained via this process determines the utility of the network path for executing entanglement enabled tasks and for the design of entanglement routing protocols in the quantum network. Here, we study the end-to-end entanglement of paths in a quantum network when the edges are characterised by multi-parameter quantum states that may be considered to be the output of arbitrary and unknown quantum channels described by the network's edges. We find that over ensembles of multi-parameter states with fixed concurrence but varying density matrix elements, the end-to-end entanglement takes a range of values upper bounded by a function of the concurrence of the network-edge states. The scaling behaviour of the average end-to-end entanglement reveals that its distribution gets increasingly concentrated around the mean as the paths become longer. For a network path of a given length, the average end-to-end entanglement vanishes for edge concurrence values below a threshold that increases with the path-length. Whereas, for edge concurrence values greater than the threshold the average end-to-end entanglement increases faster with the length of the path.  As an implication of our results, we show that the optimal path for entanglement distribution between a pair of end nodes, connected by alternate paths with multi-parameter states along the edges, can be indeterminate given only entanglement guarantees along the network edges. 

\end{abstract}

\maketitle

\section{Introduction}

\begin{figure}
    \centering
    \includegraphics[width=\linewidth]{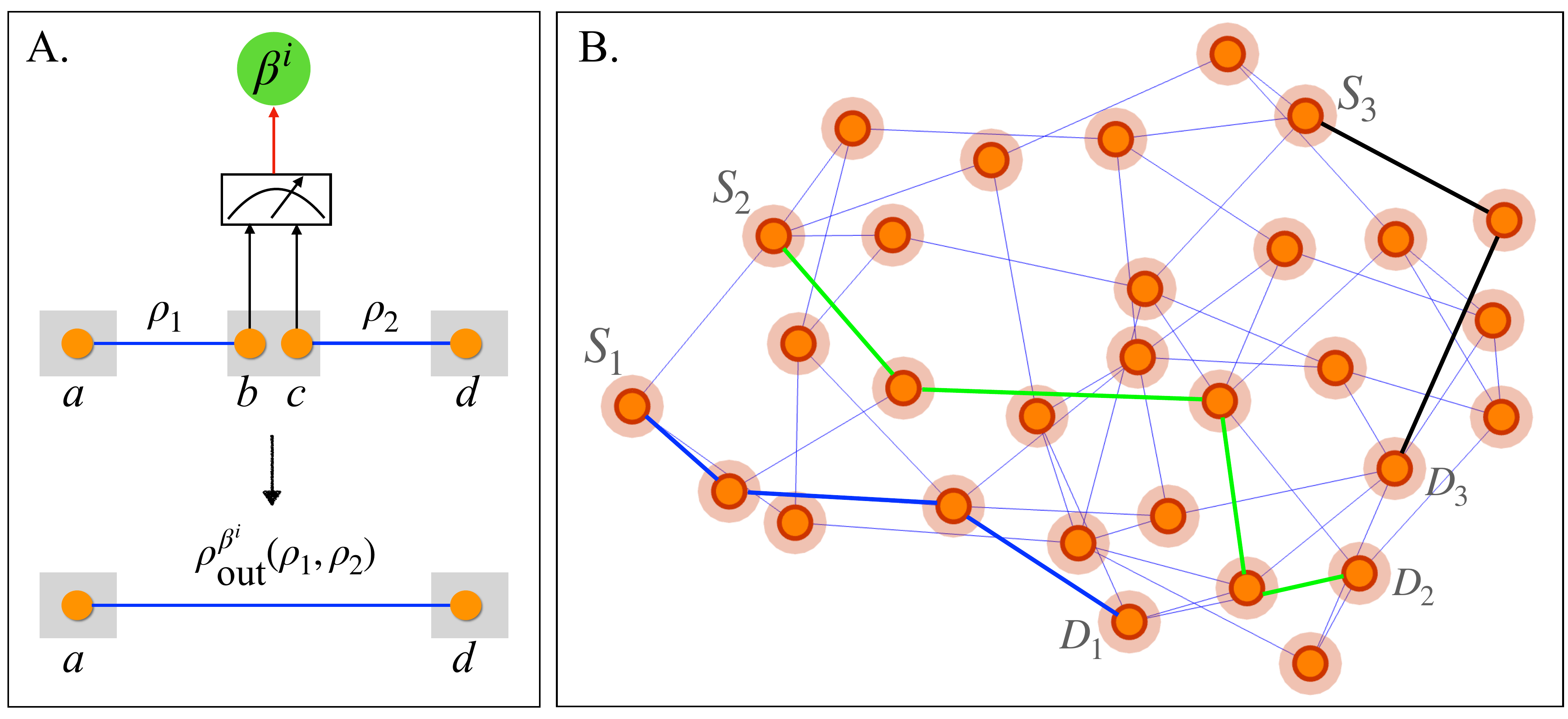}
    \caption{(Color online) (A) Illustration of entanglement swapping operation between two entangled states shared by adjacent qubit pairs $a-b$ and $c-d$. This is achieved by performing a Bell state measurement on qubits $b,c$ yielding an output state, $\rho_{\textrm{out}}^{\beta^i}(\rho_1,\rho_2)$, of qubits $a,d$ conditioned on the measurement outcome, $\beta^i,i=0,1,2,3$. (B) A random quantum network with pairs of source-destination, $S,D$, nodes connected by paths of different graph lengths, $l=2$ (Black), $l=3$ (Blue), $l=4$ (Green).}
    \label{fig:Network_Swapping}
\end{figure}

Entanglement swapping \cite{swapping_zeillinger,swapping_vedral,swapping_zeillinger2} is a crucial quantum operation in a quantum network \cite{q_net_kimble,wehner_qnet,simon_qnet} that connects short-range entangled states on the edges of the network to create long-range entangled states between a pair of end-nodes \cite{repeater_briegel,repeater_liang,ent_topography}, see Fig.~\ref{fig:Network_Swapping}{\color{red}}. These long-range end-to-end entangled states can then be used for promising quantum information tasks ranging from secure communication to distributed quantum computation \cite{teleportation,qkd_bennett,qkd_ekert,dist_q_comput_1,dist_q_comput_2}. An understanding of the behavior of the end-to-end entanglement as a function of the magnitude of short-range entanglement (the concurrence along the edges in the path) and its scaling with path length, in terms of the number of edges, is important to obtain as quantum networks mature \cite{small_net1,small_net2,small_net3,small_net4,small_net5,small_net6, satelite_ent_dist} because first, it allows one to determine if entanglement distribution along a path can satisfy the entanglement threshold requirement for a quantum task \cite{wehner_qnet,zhang2022device,zapatero2023advances}; next, it can guide the design of entanglement routing and distribution protocols \cite{ent_topography,leone2024costvectoranalysis}; and further, whether multiple network paths between the same pair of end nodes may be utilised for sophisticated entanglement distribution protocols such as multipath entanglement purification \cite{mep_conf_sohel,mep_sohel}.

In an entanglement swapping operation, measurement in an entangled basis of one-half of each state in a pair of adjacent network-edge states, can yield a set of entangled states of the unmeasured qubits of the two pairs conditioned on the outcomes of the measured qubits, Fig.~\ref{fig:Network_Swapping}{\color{red}A}. The average entanglement of the obtained post-swapped state in this process, as measured by the output average concurrence, can be calculated as the sum of the concurrences of the output states weighted by the probabilities of their outcomes \cite{avgconcurrence_santra}. Since long-range entanglement distribution in a network requires multiple entanglement swappings between the network edges, the output average concurrence obtained after these multiple swaps is a natural quantity to characterise the end-to-end entanglement - which can be determined exactly given the density matrices of the states along the edges comprising the path. This latter requirement is, however, difficult to satisfy experimentally since it requires an exact characterisation of the quantum channels \cite{exp_channel_charac,exp_charac_process} connecting the network nodes - a task whose complexity scales with the number of network nodes. Further, it may be difficult to exactly characterise physical links, such as fiber-optical \cite{fiber_qkd_time_dep} or free-space links \cite{free_space_time_dep}, that can have time-dependent quantum channel properties. Moreover, in some situations although the output quantum states depend on the channel characteristics the concurrence of the state does not \cite{brodsky_channel}. Therefore, it is important to have an estimate of the end-to-end entanglement as a function of the concurrence of the states over the edges along a network path and its length without regard to the exact form of their density matrices.

In this paper we study the end-to-end entanglement as a function of the concurrence of the states along the edges of the path
and of the path length, in the scenario, where, in the description of the quantum network the concurrence values of the states along its edges are specified but the precise form of their density matrices may not be available.  The scenario we consider relaxes the requirement for an exact characterisation of the quantum channel over the network-edges in the network description. Note that general quantum channels have output states characterised by multiple parameters in their density matrices  \cite{swapping_santra} with their concurrence being a function of these multiple parameters. Further, for a given value of concurrence there exist an arbitrary number of states with distinct density matrices which when swapped result in different values of the output post-swap concurrence. Thus, we obtain the average end-to-end entanglement and its range by sampling over ensembles of input multi-parameter states with fixed as well as varying purities consistent with the given concurrence along the edges of the path. We iterate this process for different concurrences over the entire domain of input concurrence values between zero and one. This allows us to obtain the scaling of the end-to-end entanglement and its range as a function of the input edge-concurrence and the network path-length - which we interpret as the expected behavior of the end-to-end entanglement obtained in a swap-only distribution protocol along a network path comprised of arbitrary and unknown quantum channels given only entanglement guarantees for the states produced by these channels.

We find that over the ensembles of multi-parameter states with fixed concurrence, $C$, but with distinct density matrices, the end-to-end entanglement can take a range of values upper bounded by a function of the concurrence of the network-edge states. From the scaling behaviour of the average end-to-end entanglement we find that its distribution gets increasingly concentrated around the mean as the network paths become longer. For a network path of a given length, the average end-to-end entanglement vanishes for edge concurrence values below a threshold, $C_{\textrm{Th}}(l)=(1-\xi/l),\xi=1.35$, that increases with the path-length, $l\geq 2$, measured as the number of edges along the path. On the other hand, for edge concurrence values greater than the threshold the average end-to-end entanglement increases rapidly with the length of the path and shows a scaling, $m_l(C^l-(C_{\textrm{Th}}(l))^l)$, where, $m_l$ increases with $l$.  Finally, we show that the optimal path for entanglement distribution between a pair of end nodes, connected by alternate paths with multi-parameter states along the edges, can be indeterminate given only entanglement guarantees along the network edges without any further information about the network edge states.

The structure of the paper is as follows : In Sec. \ref{section_2} we revisit the basics of entanglement swapping. In Sec. \ref{section_3} we first present the form of the analytically computable output average concurrences after swapping of different single-parameter states, then we analyze the output average concurrence post entanglement swapping of two arbitrary multi-parameter states of known input concurrences followed by similar results for entanglement swapping of one single-parameter state with a multi-parameter state. In Sec. \ref{section_4} we discuss the numerical results obtained for multiple entanglement swappings along network paths of increasing length and describe the ensembles of multi-parameter states from where the network edge states are sampled. In Sec. \ref{section_5} we show that the optimal path for entanglement distribution between a pair of network nodes with edges described by multi-parameter states cannot be determined given only the concurrence values of these states. Finally in Sec. \ref{section_6} we conclude by discussing our results and scope for future work.
\section{Entanglement Swapping Preliminaries}
\label{section_2}
Entanglement swapping operations can be used to quantum correlate two previously uncorrelated quantum systems, say $a$ and $d$, by joint entangled measurements on parts, $b$ and $c$, of two distinct entangled states $\rho_1$ on systems $a$ and $b$ and $\rho_2$ on systems $c$ and $d$, respectively, see Fig. \ref{fig:Network_Swapping}{\color{red}A}.  A complete basis of entangled states for the qubits, $b$ and $c$, can be taken to be the Bell-basis of states, $\{\ket{\beta^0},\ket{\beta^1},\ket{\beta^2},\ket{\beta^3}\}$, which in their computational basis may be expressed as,
\begin{align}
    \ket{\beta^0}&:=\ket{\Phi^+}_{bc}=\frac{1}{\sqrt{2}}(\ket{00}_{bc}+\ket{11}_{bc}),\nonumber\\
    \ket{\beta^1}&:=\ket{\Phi^-}_{bc}=\frac{1}{\sqrt{2}}(\ket{00}_{bc}-\ket{11}_{bc}),\nonumber\\
    \ket{\beta^2}&:=\ket{\Psi^+}_{bc}=\frac{1}{\sqrt{2}}(\ket{01}_{bc}+\ket{10}_{bc}),\nonumber\\
    \ket{\beta^3}&:=\ket{\Psi^-}_{bc}=\frac{1}{\sqrt{2}}(\ket{01}_{bc}-\ket{10}_{bc}).
    \label{measurement_basis}
\end{align}

Such a Bell-state measurement (BSM) of qubits $b,c$ yields a potentially entangled state of the qubits $a,d$ given by,
\begin{equation}
\rho_{\textrm{out}}^{\beta^i}(\rho_1,\rho_2)=\frac{\bra{\beta^i}(\rho_{1}\otimes \rho_2)\ket{\beta^i}}{\text{Tr}_{ad}[\bra{\beta^i}(\rho_{1}\otimes \rho_2)\ket{\beta^i}]},
    \label{swapped_state}
\end{equation}
conditioned on the measurement outcome, $\ket{\beta^i},i=0,1,2,3$, each of which occurs with a probability,
\begin{equation}
    p^{\beta^i}=\text{Tr}_{ad}[\bra{\beta^i}(\rho_{1}\otimes \rho_2)\ket{\beta^i}].
    \label{swapped_probability}
\end{equation} 

The average concurrence of the output state obtained via entanglement swapping, $\overline{C}(\rho_1,\rho_2)$, can be calculated as the concurrence of the conditional output state given in Eq. (\ref{swapped_state}) weighted by the probability of its outcome \cite{avgconcurrence_santra}. Thus,
\begin{align}
    \overline{C}:=\sum_{i} p^{\beta^i} \mathcal{C}(\rho_{\textrm{out}}^{\beta^i}),
    \label{avg_concurrence}
\end{align}
is a measure of the concurrence obtained as the output of the entanglement swapping process on average with, $\mathcal{C}(\rho)$, being the function denoting the concurrence of the state, $\rho$. Note that we have suppressed the arguments $\rho_1,\rho_2$ of the function $\overline{C}$ in the left hand side of Eq. (\ref{avg_concurrence}) for brevity of use in the rest of the paper. 

The concurrence function, $\mathcal{C}(\rho)$, of an arbitrary two-qubit density matrix $\rho$ can be evaluated as \cite{concurrence_wooters1,concurrence_wooters2},
\begin{align}
    \mathcal{C}(\rho)=\text{Max} \left[0,2\lambda_{M}-\Tr(R)\right],
    \label{concurrence}
\end{align}
where, $R=\rho(\sigma_y\otimes\sigma_y)\rho^*(\sigma_y\otimes\sigma_y)$, is a matrix obtained from the original density matrix $\rho$, its complex-conjugate $\rho^*$ and $\sigma_y=\begin{pmatrix}0 & -i\\i & 0\end{pmatrix}$ being one of the Pauli-spin matrices with $\lambda_M$ being the maximum eigenvalue of the matrix, $R$.

We term those two-qubit states, $\rho$, whose concurrence, $C(\rho)$, is a function of a lone parameter describing the state as a single-parameter state in contrast to multi-parameter states where multiple parameters appearing in the density matrix of the state appear in the expression for its concurrence. The entanglement swapping behaviour of single- and multi-parameter states is described next.

\section{Entanglement swapping of single and multi parameter states\label{section_3}}

Given the concurrence values, $C_1=\mathcal{C}(\rho_1),C_2=\mathcal{C}(\rho_2)$, of the density matrices, $\rho_1,\rho_2$, that are inputs to the entanglement swapping operation, we consider in this section the average output concurrence after a single swapping of the post-swapped state, $\overline{C}$, given by Eq. (\ref{avg_concurrence}) in terms of the inputs $C_1$ and $C_2$. Some of the results here have been obtained by other authors and we provide appropriate references in those cases. 

The main message this section conveys is that for the single-parameter class of states the average output concurrence, $\overline{C}$, can be obtained deterministically, whereas, when one or both of the input states used for entanglement swapping is of the multi-parameter class the average output concurrence can take a range of values , $0\leq \overline{C} \leq C_1C_2$, implying that the input concurrence values are not sufficient to determine $\overline{C}$ exactly.

\subsection{Entanglement swapping of single-parameter states}
\label{section_2a}
Consider the following types of two-qubit network-edge states that may be shared by the quantum systems, $a$ and $b$, shown in Fig. \ref{fig:Network_Swapping}\textcolor{red}{A},
\begin{align}
\ket{\psi(\lambda)}&=\sqrt{\lambda}\ket{00}_{ab}+\sqrt{1-\lambda}\ket{11}_{ab},\label{pure_state}\\
\rho_{W}(\gamma)&=(1-\gamma)\ket{\Psi^-}_{ab}\bra{\Psi^-}+\frac{\gamma}{4}\boldsymbol{I_4},\label{W_state}\\
\rho_{I}(\alpha)&=(1-\alpha)\ket{\Phi^+}_{ab}\bra{\Phi^+}+\frac{\alpha}{4}\boldsymbol{I_4},\label{I_state}\nonumber\\
\end{align}
which are, respectively, a pure state expressed in the Schmidt basis for the qubits \cite{nielsen_book}, a Werner state \cite{werner_state} and an Isotropic state \cite{isotropic_state} - for all of which the concurrence of the states are a function of the single parameter describing the state as follows,
\begin{align}
\mathcal{C}({\psi(\lambda)})&=2\sqrt{\lambda(1-\lambda)}\label{pure_state_conc},\\
\mathcal{C}(\rho_W(\gamma))&=\text{Max}(0,1-\frac{3}{2}\gamma),\label{werner_conc}\\
\mathcal{C}(\rho_I(\alpha))&=\text{Max}(0,1-\frac{3}{2}\alpha),\label{isometric_conc}
\end{align}
- permitting their classification into the single-parameter class of states. 

When two states of the single-parameter class are swapped, the average output concurrence, $\overline{C}$, given by Eq. (\ref{avg_concurrence}) can be determined exactly as a function of the input concurrences. In the above cases these are given by,
\begin{align}
\overline{C}(\psi(\lambda_1),\psi(\lambda_1))&=C_1C_2,\label{swapped_conc_pure}\\
\overline{C}(\rho_W(\gamma_1),\rho_W(\gamma_2))&=\text{Max}\left[0,\frac{1}{3}(C_{1}+C_{2}+2C_{1}C_{2}-1),\right]\label{swapped_conc_werner}\\
\overline{C}(\rho_I(\alpha_1),\rho_I(\alpha_2))&=\text{Max}\left[0,\frac{1}{3}(C_{1}+C_{2}+2C_{1}C_{2}-1).\right]
\label{swapped_conc_isotropic}
\end{align}
where the expressions of $C_1, C_2$ in Eq. (\ref{swapped_conc_pure}), (\ref{swapped_conc_werner}), (\ref{swapped_conc_isotropic}) are given by Eq. (\ref{pure_state_conc}), (\ref{werner_conc}), (\ref{isometric_conc}) respectively. This summarises results obtained in references \cite{ent_topography,avgconcurrence_santra,swapping_santra}. 
\subsection{Entanglement swapping of multi-parameter states}
\label{section_2b}

\begin{figure}
    \centering
    \includegraphics[width=\linewidth]{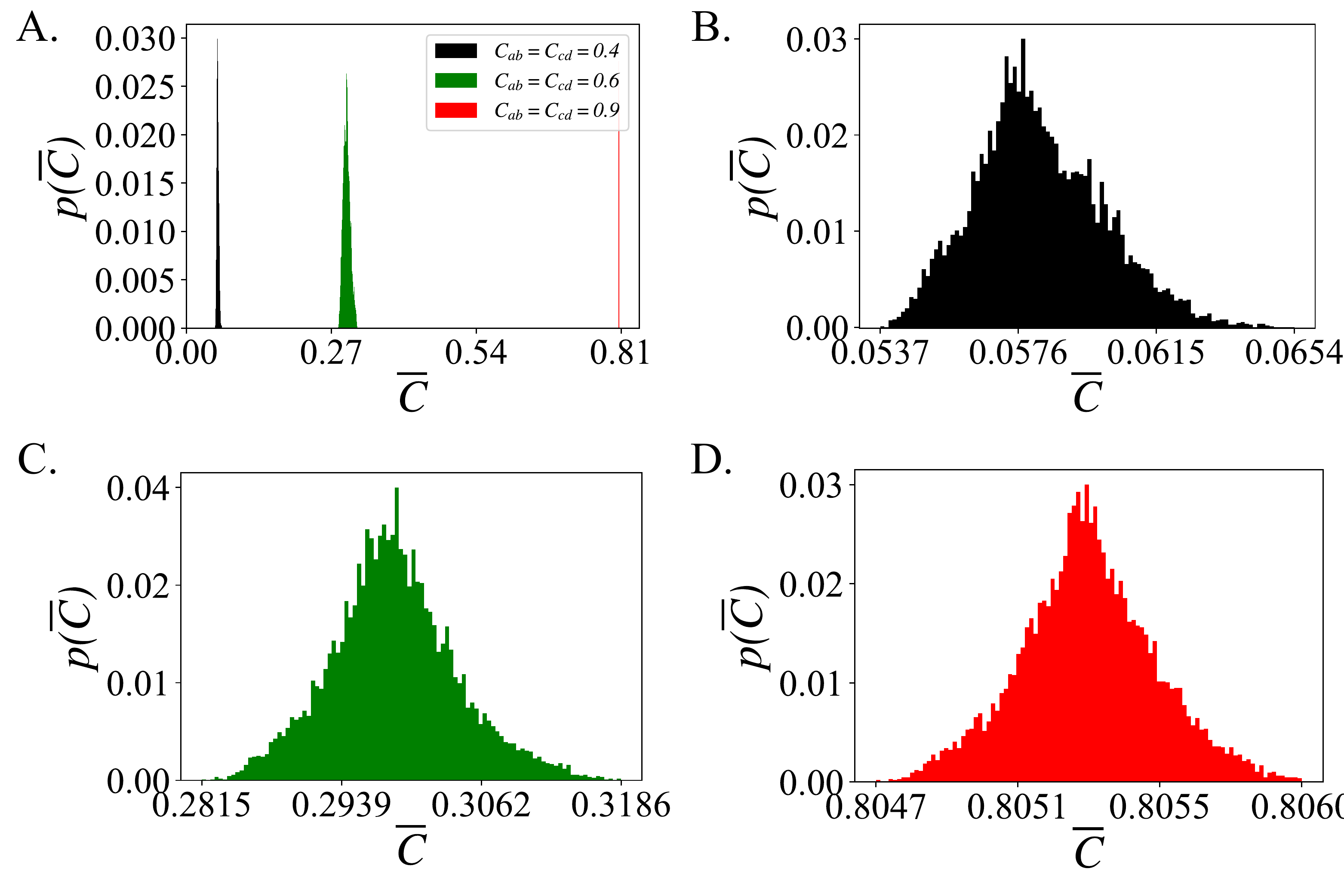}
    \caption{Probability distribution function of the output average concurrence, $\overline{C}$, after entanglement swapping of two rank-4, multiparameter states with equal concurrences, $C_1=C_2=C$, obtained by sampling over $10^4$ pairs of input states. Panel (A) shows the distributions for $C=0.4$ (black), $C=0.6$ (green) and $C=0.9$ (red) using the same scale for the axes. Panels (B) , (C) and (D) are the close-up views using different scales of the axes for clarity.}
    \label{fig:Distributions}
\end{figure}

Next, consider the following types of two-qubit network-edge states that may be shared by the quantum systems, $a$ and $b$, shown in Fig. \ref{fig:Network_Swapping}\textcolor{red}{A},
\begin{align}
\rho_{BD}(\{\lambda_i\}_i)&=\lambda_1\ket{\Phi^+}_{ab}\bra{\Phi^+}+\lambda_2\ket{\Phi^-}_{ab}\bra{\Phi^-}\nonumber\\
&+\lambda_3\ket{\Psi^+}_{ab}\bra{\Psi^+}+\lambda_4\ket{\Psi^-}_{ab}\bra{\Psi^-},\label{bd_state}\\
\rho_{X}(\{\gamma_{ij}\}_{i,j})&=\begin{pmatrix}
        \gamma_{11} & 0 & 0 & \gamma_{14}\\
        0 & \gamma_{22} & \gamma_{23} & 0\\
        0 & \gamma_{23} & \gamma_{33} & 0\\
        \gamma_{14} & 0 & 0 & \gamma_{44}
    \end{pmatrix},\label{x_state}\\
\rho(\{\alpha_{ij}\}_{i,j})&=\begin{pmatrix}
        \alpha_{11} & \alpha_{12} & \alpha_{13} & \alpha_{14}\\
        \alpha_{12} & \alpha_{22} & \alpha_{23} & \alpha_{24}\\
        \alpha_{13} & \alpha_{23} & \alpha_{33} & \alpha_{34}\\
        \alpha_{14} & \alpha_{24} & \alpha_{34} & \alpha_{44}
    \end{pmatrix},\label{mixed_state}
\end{align}
which are, respectively, a Bell-diagonal state, a so-called X-state because of the shape in which the matrix elements are arranged, and a general two-qubit density matrix, assuming that the real matrix elements (for simplicity) satisfy the requirements of a valid density operator, such as, $\Tr(\rho)=1,\rho^\mathrm{T}=\rho$. For all such states the concurrences are a function of mutiple parameters describing the state with,
 \begin{align}
 \mathcal{C}(\rho_{BD})&=\textrm{Max}\left[0,|\lambda_1-\lambda_2|-(\lambda_3+\lambda_4),|\lambda_3-\lambda_4|-(\lambda_1+\lambda_2)\right],\\
 \mathcal{C}(\rho_X)&=2~\textrm{Max}\left[0,|\gamma_{14}|-\sqrt{\gamma_{22}\gamma_{33}},|\gamma_{23}|-\sqrt{\gamma_{11}\gamma_{44}}\right],
 \label{eq:mul_conc}
 \end{align}
while for the general density matrix, $\mathcal{C}(\rho)$, can be calculated using Eq. (\ref{concurrence}).

When two states of the multi-parameter class are swapped, the average output concurrence, $\overline{C}$, given by Eq. (\ref{avg_concurrence}) cannot be obtained exactly as a deterministic function of the input concurrences, $C_1,C_2$. This is due to the conditional expressions for the concurrences of multi-parameter states, Eq. (\ref{concurrence}), (\ref{eq:mul_conc}), whose values depend on the relative magnitudes of the various density matrix elements.
Note the contrast with the results of the output concurrence in the single-parameter case shown in Eqs. (\ref{swapped_conc_pure})-(\ref{swapped_conc_isotropic}).

In fact, the output average concurrence, $\overline{C}$, after swapping multi-parameter states can take a range of values for a given pair of input concurrences, $C_1,C_2$, 
\begin{align}
0\leq\overline{C}\leq C_1C_2
\end{align}
with an upper bound given by, $C_1C_2$, as shown in Appendix (\ref{sec:ub_avg_conc}), while the (trivial) lower bound of, $\overline{C}$, is zero. It may be possible to obtain a different (non-trivial) lower bound for, $\overline{C}$, for multi-parameter states with fewer matrix elements, such as the Bell-diagonal and $X$-states shown in Eqs. (\ref{bd_state},\ref{x_state}), but we have been unable to obtain such lower bounds and mention this as an open problem.

Nevertheless, the probability distribution function of the average output concurrence, $p(\overline{C})$, depends on the sampled ensemble of input states, $\mathcal{E}_{C_1}(\rho_1)$ and $\mathcal{E}_{C_2}(\rho_2)$, with the given concurrences, $C_1$ and $C_2$. Fig. \ref{fig:Distributions} shows examples of skewed-Gaussian type of distributions obtained for, $\overline{C}$, when two general density matrices with the same concurrence, $C_1=C_2=C$, are sampled from an ensemble of states obtained using random bilateral single-qubit unitaries sampled with the Haar measure. Noteworthy, from our numerical analysis of various ensembles, $\mathcal{E}_C$, with different values of, $C$, is that the support of the distribution, $p(\overline{C})$, becomes narrow as the input concurrence, $C\to 1$ and $C\to 0$, whereas, for intermediate values, $C\approx 0.5$, the distribution, $p(\overline{C})$, is relatively wide. Further, for higher values of the input concurrence, $C\to1$, the distribution scales as, $\overline{C}\sim C^2$. In Sec. \ref{section_4} we will show that the scaling of the average output concurrence generalizes for network paths of length, $l$, where $(l-1)$ entanglement swappings are required for an end-to-end entanglement connection, in which case the average output concurrence scales as, $\overline{C}\sim C^l$.

\subsection{Entanglement swapping of a single-parameter state with a multi-parameter state\label{pure_mix}}
\label{section_2c}

Finally, consider the situation when entanglement swapping is performed between a single-parameter state with concurrence $C_1$ as one of the inputs, with the other input being a multi-parameter state with concurrence, $C_2$. In this scenario, when the single parameter state has a rank greater than one, such as the Werner or Isotropic states given by Eq. (\ref{W_state},\ref{I_state}) , the average output concurrence, $\overline{C}$, cannot be determined as a function of the input concurrences, $C_1,C_2$. In this case, similar to the situation when two multi-parameter states are swapped as described in subsec. \ref{section_2b}, one obtains a range of values for the average output concurrence with, $0\leq \overline{C}\leq C_1C_2$. However, a single-parameter state of the pure state type, Eq. (\ref{pure_state}), swapped with a multi-parameter state yields an average output concurrence given by the product of the two concurrences with $\overline{C}=C_1C_2$. We show this analytically when the multi-parameter state is of the $X$-state form, Eq. (\ref{x_state}), and numerically when the multi-parameter state is a general two-qubit density matrix, Eq. (\ref{mixed_state}) in Appendix (\ref{appendix_b}) confirming numerically obtained results obtained recently in \cite{swapping_cong}.


\section{Average output concurrence for ensembles of fixed concurrence states}
\label{section_4}
\subsection{Ensembles of states used for entanglement swapping}
To study the average output concurrence and its range for an ensemble of multi-parameter states used as inputs for entanglement swapping, we generate two distinct types of ensembles which preserve the concurrence, $C$, of the input states. 

The first type of ensemble, $\mathcal{E}_C(\rho)$ with $\mathcal{C}(\rho)=C$, consists of density matrices obtained from a two-qubit density matrix, $\rho$, using random bilateral single-qubit unitary rotations, $u_a,u_b$, of the two qubits - with the single-qubit unitaries sampled from the set of Haar random unitaries, $\textrm{U}_{\text{Haar}}(2)$, and is described by the set of states,
\begin{align}
    \mathcal{E}_C(\rho) &= \{(u_a\otimes u_b) \rho (u_a \otimes u_b)^\dagger \,|\, u_a,u_b\in \textrm{U}_{\text{Haar}}(2)\}.
    \label{ensemble_e}
\end{align}
This ensemble, $\mathcal{E}_C(\rho)$, consists of two-qubit density matrices with the same rank, purity ($\Tr(\rho^2)$) and concurrence as $\rho$. We choose $20$ different single-qubit unitaries, $u_a,u_b\in \textrm{U}(2)$, for a total of $400$ samples in the set $\mathcal{E}_C(\rho)$. The method for generating set of Haar random single qubit unitaries is described in Appendix (\ref{appendix_c1}).

The second type of ensemble, $\mathcal{S}_C(\rho)$, consists of a finite number, $N$, of full-rank two-qubit density matrices, $\rho_i,i=1,...,N$, all with the same value of their concurrence, $\mathcal{C}(\rho_i)=C$, but which are not related to each other via bilateral unitary rotations, that is,
\begin{align}
    \mathcal{S}_C &= \{\rho_i,i=1,...,N~|~\mathcal{C}(\rho_i)=C\},\nonumber\\
    \rho_k\neq (u_a\otimes u_b) \rho_i & (u_a \otimes u_b)^\dagger, u_a,u_b\in \textrm{U}_{\text{Haar}}(2) \forall \rho_i,\rho_k\in \mathcal{S}_C.
    \label{ensemble_s}
\end{align}
The ensemble $\mathcal{S}_C$ consists of two-qubit density matrices all with the same rank and concurrence - but with different purities, $(1/4)\leq \Tr(\rho^2)\leq 1$, that is, over the full range of allowed purities for two-qubit density matrices. We numerically generate the ensembles, $\mathcal{S}_C$, with, $N=10$, density matrices of different purities in each ensemble at a given value of $C$.

The motivation for considering the two ensembles, $\mathcal{E}_C(\rho)$ and $\mathcal{S}_C$, comes from the fact that, a multi-parameter state $\rho$, can be fully determined given the quantities, $\Tr(\rho^n),n=1,2,...,\infty$, with, $\Tr(\rho)=1$ (unit trace condition for valid density matrices). The ensembles, $\mathcal{E}_C(\rho)$ and $\mathcal{S}_{C}$, with fixed purities of states in the first ensemble and varying purities of the states in the second ensemble allow us to study the average output concurrence for input density matrices specified up to the first non-trivial trace of a power of their density matrix, $\Tr(\rho^2)$. 

\subsection{Case of a single swap - Range and ensemble mean of the average output concurrence}
\label{sec_singleswap}

First, we generate the ensembles, $\mathcal{E}_C(\rho_1)$ and $\mathcal{E}_C(\rho_2)$, by randomly choosing two fiducial states, $\rho_1,\rho_2\in\mathcal{S}_C$. Then, we numerically calculate the average output concurrences, $\overline{C}(\rho_i,\rho_j)$, by sampling pairs of input states from the two ensembles, $\rho_i\in \mathcal{E}_C(\rho_1)$ and $\rho_j\in \mathcal{E}_C(\rho_2)$. This gives us a list of length, $|\mathcal{E}_C(\rho_1)|\times |\mathcal{E}_C(\rho_2)|$, with values of $\overline{C}(\rho_i,\rho_j)$ for different pairs of input states. Using this list, we calculate the range of the values obtained, $\delta \overline{C}_{\mathcal{E}_C(\rho_1),\mathcal{E}_C(\rho_2)}$, as the difference between the largest and smallest average output concurrence obtained by sampling over the two  ensembles, that is,
\begin{align}
\delta \overline{C}_{\mathcal{E}_C(\rho_1),\mathcal{E}_C(\rho_2)}:=\underset{\substack{{\rho_i\in\mathcal{E}_C(\rho_1)}\\{ \rho_j\in\mathcal{E}_C(\rho_2)}}}{\max}\overline{C}(\rho_i,\rho_j)-\underset{\substack{{\rho_i\in\mathcal{E}_C(\rho_1)}\\{ \rho_j\in\mathcal{E}_C(\rho_2)}}}{\min}\overline{C}(\rho_i,\rho_j),
\label{eq:av_out_conc_range}
\end{align}
which is a random variable taking values, $0\leq  \delta \overline{C}_{\mathcal{E}_C(\rho_1),\mathcal{E}_C(\rho_2)}\leq C^2$, for each pair of ensembles, $\mathcal{E}_C(\rho_1),\mathcal{E}_C(\rho_2)$. Note that there are, $|\mathcal{S}_C|\times |\mathcal{S}_C|=N^2$, such ensemble pairs. The range, $\delta \overline{C}_{\mathcal{E}_C(\rho_1),\mathcal{E}_C(\rho_2)}$, is thus the sample spread of the average output concurrence, for a given ensemble pair, obtained upon swapping two states of fixed concurrence, $C$, but with the same purity (when $\rho_1=\rho_2$) or different purities (when $\rho_1\neq\rho_2$).

Further, over the pair of ensembles, $\mathcal{E}_C(\rho_1)$ and $\mathcal{E}_C(\rho_2)$, we obtain the mean average output concurrence, $\langle\overline{C}\rangle_{\mathcal{E}_C(\rho_1),\mathcal{E}_C(\rho_2)}$, by numerically obtaining the probability distribution function, $p(\overline{C})$, with, 
\begin{align}
\overline{C}\in[\underset{\substack{{\rho_i\in\mathcal{E}_C(\rho_1)}\\{ \rho_j\in\mathcal{E}_C(\rho_2)}}}{\max}\overline{C}(\rho_i,\rho_j),\underset{\substack{{\rho_i\in\mathcal{E}_C(\rho_1)}\\{ \rho_j\in\mathcal{E}_C(\rho_2)}}}{\min}\overline{C}(\rho_i,\rho_j)]
\end{align}
and calculating,
\begin{align}
\langle\overline{C}\rangle_{\mathcal{E}_C(\rho_1),\mathcal{E}_C(\rho_2)}=\int\limits_{\underset{\substack{{\rho_i\in\mathcal{E}_C(\rho_1)}\\{ \rho_j\in\mathcal{E}_C(\rho_2)}}}{\min}\overline{C}(\rho_i,\rho_j)}^{\underset{\substack{{\rho_i\in\mathcal{E}_C(\rho_1)}\\{ \rho_j\in\mathcal{E}_C(\rho_2)}}}{\max}\overline{C}(\rho_i,\rho_j)}d\overline{C}~p(\overline{C})\overline{C}.
\label{eq:av_out_conc_mean}
\end{align}
Note, that the quantity, $0\leq \langle\overline{C}\rangle_{\mathcal{E}_C(\rho_1),\mathcal{E}_C(\rho_2)}\leq C^2$, is also a random variable taking different values depending on the pair, $\rho_1,\rho_2\in \mathcal{S}_C$, of sampled fiducial states with there being $N^2$ such pairs.

Second, we sample pairs of fiducial states, $\rho_i,\rho_j$, over the ensemble, $\mathcal{S}_C$, and numerically calculate the average range, $\langle\delta\overline{C}\rangle_{\mathcal{S}_C}$, via,
\begin{align}
\langle\delta\overline{C}\rangle_{\mathcal{S}_C}:=\frac{\sum\limits_{\rho_i,\rho_j\in \mathcal{S}_C}\delta\overline{C}_{\mathcal{E}_C(\rho_i),\mathcal{E}_C(\rho_j)}}{|\mathcal{S}_C|^2},
\label{eq:average_range}
\end{align}
where, $\delta\overline{C}_{\mathcal{E}_C(\rho_i),\mathcal{E}_C(\rho_j)}$, is given by Eq. (\ref{eq:av_out_conc_range}).

Finally, using the values of the ensemble means, $\langle\overline{C}\rangle_{\mathcal{E}_C(\rho_1),\mathcal{E}_C(\rho_2)}$ given by Eq. (\ref{eq:av_out_conc_mean}), we obtain the average ensemble mean, $\langle\langle\overline{C}\rangle\rangle_{\mathcal{S}_C}$, when $\rho_1,\rho_2$ are sampled from $\mathcal{S}_C$, as,
\begin{align}
\langle\langle\overline{C}\rangle\rangle_{\mathcal{S}_C}:=\frac{\sum\limits_{\rho_i,\rho_j\in \mathcal{S}_C}\langle\overline{C}\rangle_{\mathcal{E}_C(\rho_i),\mathcal{E}_C(\rho_j)}}{|\mathcal{S}_C|^2}.
\label{eq:stat_av_ensemble_mean}
\end{align}

\begin{figure}
    \centering
        \includegraphics[width=.9\columnwidth]{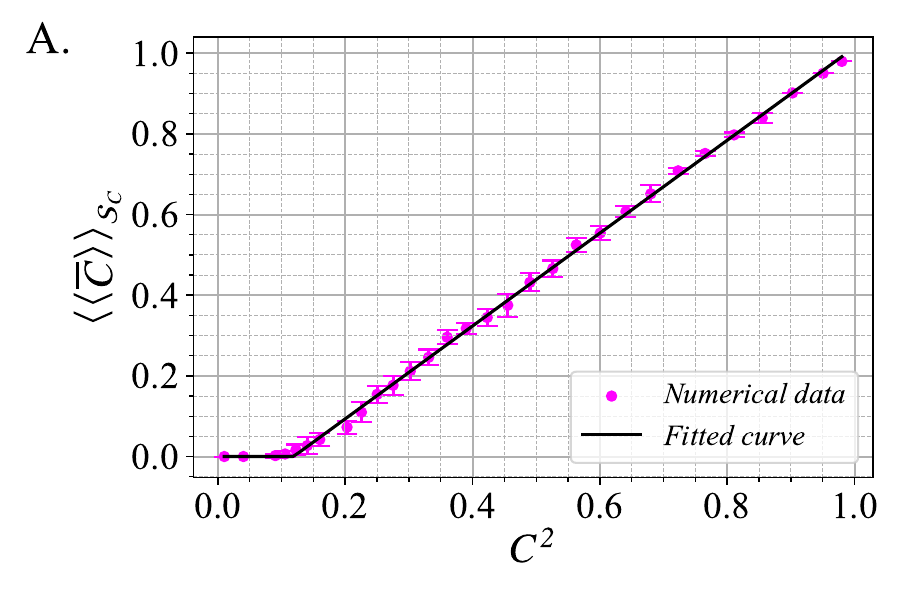}\\
    \includegraphics[width=.9\columnwidth]{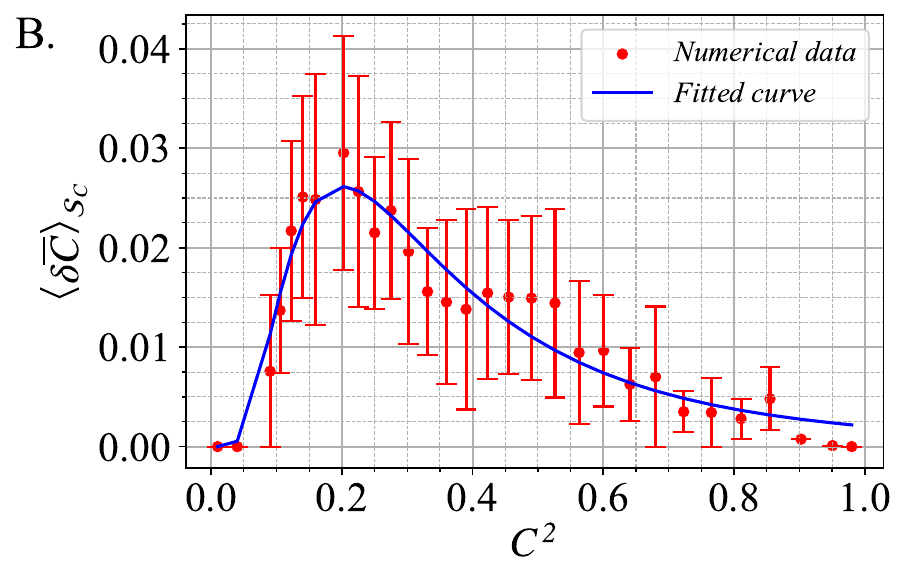}
    \caption{(Color online) (A) Statistical average of the ensemble means of the average output concurrence, 
    $\langle\langle\overline{C}\rangle\rangle_{\mathcal{S}_C}$, vs. the product of the input concurrences, $C^2$, after swapping two rank-4, multi-parameter states of the same concurrences, $C_1=C_2=C$ sampled from the ensembles, $\mathcal{E}_C(\rho_1),\mathcal{E}_C(\rho_2)$ with all possible choices of $\rho_1,\rho_2\in\mathcal{S}_C$. Data points are shown in Pink while the vertical bars represent the statistical standard deviation. The black solid line represents the least squares fit of the data points according to Eq.~(\ref{two_swap_mean}). (B) Statistical average of the range, $\langle\delta\overline{C}\rangle_{\mathcal{S}_C}$, of the average output concurrence vs. the product of the input concurrences, $C^2$, over the same ensembles as above shown in Red dots with the standard deviation shown with the red vertical bars. The Blue solid line represents a least squares fit for the data points.}
    \label{fig:Swapping_Plots}
\end{figure}

The two quantities, $\langle\langle\overline{C}\rangle\rangle_{\mathcal{S}_C}$ given by Eq. (\ref{eq:stat_av_ensemble_mean}) and $\langle\delta\overline{C}\rangle_{\mathcal{S}_C}$ given by Eq. (\ref{eq:average_range}), which are the average ensemble mean and the average range, respectively - convey valuable information about the post-swap concurrence obtained for ensembles of rank-4, multi-parameter states with input concurrences, $C\in[0,1]$, and a range of purities, $(1/4)\leq\Tr(\rho^2)\leq 1$. The average of the ensemble means, $\langle\langle\overline{C}\rangle\rangle_{\mathcal{S}_C}$, yields the average output concurrence obtained, whereas, the average ensemble range, $\langle\delta\overline{C}\rangle_{\mathcal{S}_C}$, reveals the range of values around the average ensemble mean that may be obtained when swapping multi-parameter mixed states of a given concurrence without requiring knowledge of the form of their density matrix.

Plotting, first, the statistical average of the ensemble means, $\langle\langle\overline{C}\rangle\rangle_{\mathcal{S}_C}$,  in Fig.~\ref{fig:Swapping_Plots}\textcolor{red}{A}, with respect to the input concurrence, $C$, we find that it is zero below a threshold value of the input concurrence, $C_{\textrm{Th}}$, and thereafter increases linearly as the square of the input concurrence, that is,
\begin{align}
\langle\langle\overline{C}\rangle\rangle_{\mathcal{S}_C}=\text{max}\left[0, m(C^2-C^2_{\textrm{Th}})\right].
    \label{two_swap_mean}
\end{align}
with the values, $C_{\textrm{Th}}\approx0.34$ and $m\approx 1.15$, obtained using numerical fitting of the data. Note, however, that when both input concurrences have the maximal value, $C=1$, pertaining to the pure-state input situation, then Eq. (\ref{swapped_conc_pure}) implies that, $\langle\langle\overline{C}\rangle\rangle_{\mathcal{S}_C}=1$, and therefore, $m(C^2-C^2_{\textrm{Th}})|_{C=1}=1$, giving us,
\begin{align}
C_{\textrm{Th}}=\sqrt{1-\frac{1}{m}},
\end{align}
which shows that the threshold input concurrence for non-zero average ensemble mean and the slope of the linear part are related to each other. Using the numerical value, $m=1.15$, we obtain, $C_{\textrm{Th}}=0.36$, that closely matches the value obtained using a numerical data fit.

Next, plotting the average range, $\langle\delta\overline{C}\rangle_{\mathcal{S}_C}$ with respect to the input concurrence $C$, in Fig.~\ref{fig:Swapping_Plots}\textcolor{red}{B}, we find that its magnitude is small relative to the value $\langle\langle\overline{C}\rangle\rangle_{\mathcal{S}_C}$ for higher values of the input concurrence, $C\to 1$. However, for a wide range of values, $.36\lesssim C\lesssim 0.7$, it can be significant with a value of $10\%$ or more. Further, we see that the ratio $\langle\delta\overline{C}\rangle_{\mathcal{S}_C}/\langle\langle\overline{C}\rangle\rangle_{\mathcal{S}_C}$, takes its maximum value close to the threshold value of the concurrence, $C\approx.36$, as shown in Fig. \ref{fig:relative_range}. This implies that ensembles of states with lower values of concurrence can have a wide range of average output concurrences, $\overline{C}$.

\subsection{Case of multiple swaps - Range and ensemble mean of the average output concurrence for network paths\label{multiple_swap_mixed}}

Multiple entanglement swapping operations are required for end-to-end entanglement distribution over quantum network paths comprised of 3 or more edges, See Fig. \ref{fig:Network_Swapping}. This motivates the generalisation of the numerical analysis of the subsection above to obtain the statistical average of the ensemble means, $\langle\langle\overline{C}_l\rangle\rangle_{\mathcal{S}_C}$, and the average range, $\langle\delta\overline{C}_l\rangle_{\mathcal{S}_C}$, to the case of $(l-1)$-swaps, where, $l$ is the number of edges in the path. 

The average output concurrence, $\overline{C}_l$, after entanglement swapping of states $\rho_1, \dots, \rho_l$ distributed along $l$ edges of a path - which we interpret as a measure of the end-to-end entanglement of the network path - can be obtained by generalizing the expression for the case of a single swap, Eq.~(\ref{avg_concurrence}) from Sec. \ref{section_2}, as,
\begin{align}
\overline{C}_l&=\sum_{i_1=0}^{3} \sum_{i_2=0}^3 \dots \sum_{i_{l-1}=0}^3~p^{\beta^{i_1}}p^{\beta^{i_2}}\dots p^{\beta^{i_{l-1}}}\nonumber\\
    &\mathcal{C}[\rho_\text{out}^{\beta^{i_{l-1}}}(\rho_l,\rho_\text{out}^{\beta^{i_{l-2}}}(\rho_{l-1},\rho_\text{out}^{\beta^{i_{l-3}}}(\dots(\rho_3,\rho_\text{out}^{\beta^{i_1}}(\rho_2,\rho_1))\dots)))],
    \label{av_out_conc_l}
\end{align}
where, $\beta^{i_k}$ for $k=1,\dots,(l-1)$ is the outcome of the Bell-state measurement performed at $k^\text{th}$ intermediate node along the network path. The probability of the BSM outcome $\beta^{i_k}$ at the $k^\textrm{th}$ intermediate node is represented by $p^{\beta^ {i_k}}$ and the function $\mathcal{C}[.]$ is the concurrence of the state in its argument. The right hand side of Eq. (\ref{av_out_conc_l}) is a sum of over $4^{(l-1)}$ terms with each term being the concurrence of the end-to-end state obtained after $(l-1)$ swaps multiplied by the probability of obtaining the state.  Note that when all the $l$-states along the edges of the path are pure states with concurrence, $C$, we obtain that, $\overline{C}_l=C^l$, whereas, if the states are multi-parameter states then, $0\leq \overline{C}_l\leq C^l$.

To study the scaling of the average output concurrence, $\overline{C}_l$, with the input concurrence, $C$, for multiple swaps of multi-parameter states along the edges of the path we resort to the generalisation of the ensemble approach that we used for the case of a single swap described in Sec. \ref{sec_singleswap}. In this case, first, we numerically evaluate the average output concurrence, $\overline{C}_l(\rho_{i_1},\dots,\rho_{i_l})$ using Eq. (\ref{av_out_conc_l}), by sampling sets of $l$ input states independently chosen from $l$ ensembles, $\rho_{i_1}\in\mathcal{E}_C(\rho_1),\dots,\rho_{i_l}\in\mathcal{E}_C(\rho_l)$. Thus, we obtain a list of values for, $\overline{C}_l(\rho_{i_1},\dots,\rho_{i_l})$, with the size of the list corresponding to the number of input sets of states given by, $|\mathcal{E}_C(\rho_1)|\times\dots\times |\mathcal{E}_C(\rho_l)|$, where the ensemble $\mathcal{E}_C(\rho_{i})$ is obtained by Haar random local unitary rotations of states as described in Eq. (\ref{ensemble_e}). The range of the values in this list is given by,
\begin{align}
&\delta \overline{C}_{l|\mathcal{E}_C(\rho_1),\dots,\mathcal{E}_C(\rho_l)}:=\nonumber\\
&\underset{\substack{{\rho_{i_1}\in\mathcal{E}_C(\rho_1)}\\ \vdots\\{ \rho_{i_l}\in\mathcal{E}_C(\rho_l)}}}{\max}\overline{C_l}(\rho_{i_1},\dots,\rho_{i_l})-\underset{\substack{{\rho_{i_1}\in\mathcal{E}_C(\rho_1)}\\ \vdots\\{ \rho_{i_l}\in\mathcal{E}_C(\rho_l)}}}{\min}\overline{C_l}(\rho_{i_1},\dots,\rho_{i_l}),
\label{eq:av_out_path_conc_range}
\end{align}
which generalizes the range of the output concurrence for the case of a single swap given by Eq. (\ref{eq:av_out_conc_range}). Note that the range given by Eq. (\ref{eq:av_out_path_conc_range}) is a random variable taking values, $0\leq \delta \overline{C}_{l|\mathcal{E}_C(\rho_1),\dots,\mathcal{E}_C(\rho_l)}\leq C^l$, for each set of ensembles, $\mathcal{E}_C(\rho_1),\dots,\mathcal{E}_C(\rho_l)$, with there being, $|\mathcal{S}_C|^l=N^l$, such sets of ensembles, one for each choice of the fiducial states, $\rho_1,\rho_2,\dots,\rho_l\in \mathcal{S}_C\times\mathcal{S}_C\times\dots\times\mathcal{S}_C$, that generate the ensembles, $\mathcal{E}_C(\rho_i)$.

Next, for each set $\mathcal{E}_C(\rho_1),\dots,\mathcal{E}_C(\rho_l)$ of ensembles among the $N^l$ possible sets, we obtain the ensemble mean of the average output concurrence, $\langle\overline{C}_l\rangle_{\mathcal{E}_C(\rho_1),\dots,\mathcal{E}_C(\rho_l)}$,
by averaging over the list of values of $\overline{C}_l(\rho_{i_1},\dots,\rho_{i_l})$ obtained above  as,
\begin{align}
\langle\overline{C}_l\rangle_{\mathcal{E}_C(\rho_1),\dots,\mathcal{E}_C(\rho_l)}=\int\limits_{\underset{\substack{{\rho_{i_1}\in\mathcal{E}_C(\rho_1)}\\\vdots\\{ \rho_{i_l}\in\mathcal{E}_C(\rho_l)}}}{\min}\overline{C}_l(\rho_{i_1},\dots,\rho_{i_l})}^{\underset{\substack{{\rho_{i_1}\in\mathcal{E}_C(\rho_1)}\\\vdots\\{ \rho_{i_l}\in\mathcal{E}_C(\rho_l)}}}{\max}\overline{C}_l(\rho_{i_1},\dots,\rho_{i_l})}d\overline{C}_l~p(\overline{C}_l)\overline{C}_l.
\label{eq:av_out_path_conc_mean}
\end{align}
which generalizes the ensemble mean for the case of a single-swap given by Eq.~(\ref{eq:av_out_conc_mean}) to the case of multiple swaps. Note, that the ensemble mean given by Eq. (\ref{eq:av_out_path_conc_mean}), is also a random variable taking different values, $0\leq \langle\overline{C}_l\rangle_{\mathcal{E}_C(\rho_1),\dots,\mathcal{E}_C(\rho_l)}\leq C^l$, depending on the set of fiducial states, $\rho_1,\dots,\rho_l\in \mathcal{S}_C\times\mathcal{S}_C\dots\times\mathcal{S}_C$, of sampled fiducial states with there being $N^l$ such sets.

\begin{figure}
    \centering
    \includegraphics[width=.9\columnwidth]{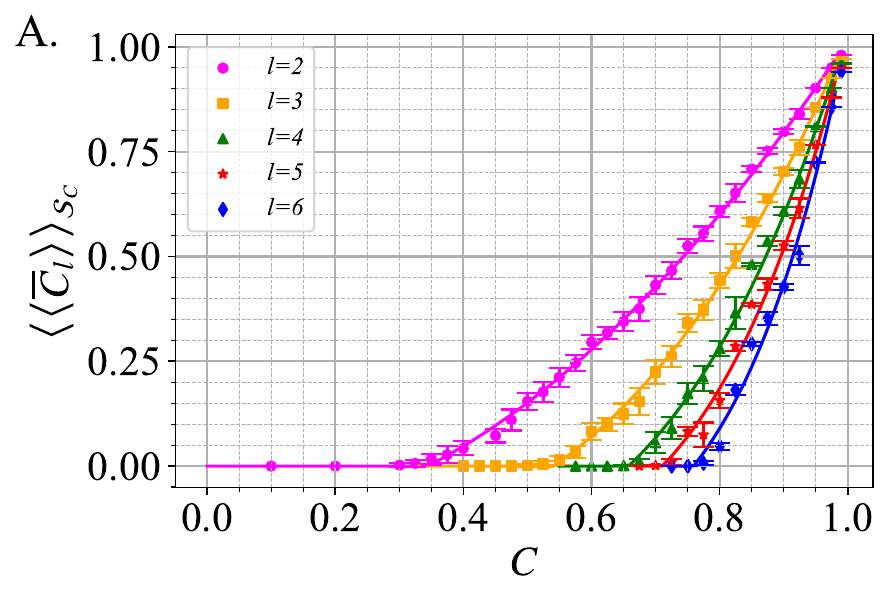}\\
    \includegraphics[width=.9\columnwidth]{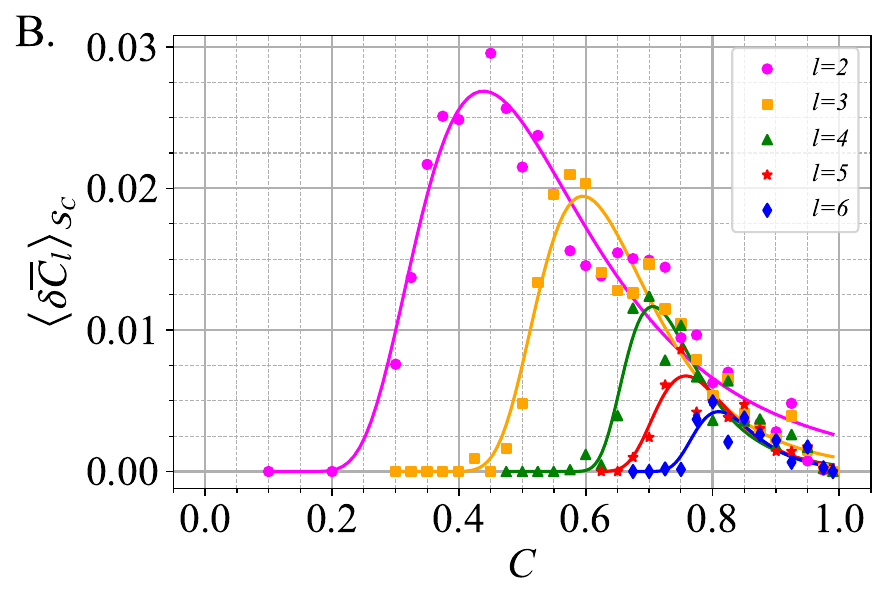}
    \caption{(A) Statistical average of the ensemble means of the average output concurrence, $\langle\langle \overline{C}_l\rangle \rangle_{\mathcal{S}_C}$, vs. the concurrence, $C$, of multi-parameter states along paths of different lengths obtained by sampling over the ensembles, $\mathcal{E}_C(\rho_1),\mathcal{E}_C(\rho_2)$ with all possible choices of $\rho_1,\rho_2\in\mathcal{S}_C$. The shapes of the data points correspond to the path length, $l=2,3,4,5,6$, shown in the legend. The solid lines represent the least squares fit for the data points according to Eq.~(\ref{net_swap_mean}). (B) Statistical average of the range of the average output concurrence, $\langle\delta \overline{C}_l\rangle_{\mathcal{S}_C}$, vs. the concurrence of the multi-parameter states along the path, $C$, for different path lengths, $l$, obtained by sampling over the same ensembles as above. The data points of different shapes correspond to the path lengths as above and the solid lines represent the least squares fit.}
    \label{fig:Swapping_Network}
\end{figure}

Finally, we sample set of $l$ fiducial states $\rho_{i_1},\dots,\rho_{i_l}$ from the ensemble $\mathcal{S}_C$ and numerically obtain the average range as,
\begin{align}
\langle\delta\overline{C}_l\rangle_{\mathcal{S}_C}:=\frac{\sum\limits_{\rho_{i_1},\dots,\rho_{i_l}\in \mathcal{S}_C}\delta\overline{C}_{l|\mathcal{E}_C(\rho_{i_1}),\dots,\mathcal{E}_C(\rho_{i_l})}}{|\mathcal{S}_C|^l},
\label{eq:average_path_range}
\end{align}
and calculate the statistical average of the ensemble means via,
\begin{align}
\langle\langle\overline{C}_l\rangle\rangle_{\mathcal{S}_C}:=\frac{\sum\limits_{\rho_{i_1},\dots,\rho_{i_l}\in \mathcal{S}_C}\langle\overline{C}_l\rangle_{\mathcal{E}_C(\rho_{i_1}),\dots,\mathcal{E}_C(\rho_{i_l})}}{|\mathcal{S}_C|^l},
\label{eq:stat_av_ensemble_mean_l}
\end{align}
which generalise similar quantities for the case of single-swap given by Eqs. (\ref{eq:average_range}) and (\ref{eq:stat_av_ensemble_mean}). 

Plotting the average of the ensemble means, $\langle\langle\overline{C}_l\rangle\rangle_{\mathcal{S}_C}$, with respect to the input concurrence, $C$, in Fig.~\ref{fig:Swapping_Network}\textcolor{red}{A} for paths of different length, $l=2,3,4,5,6$, we find that it may be expressed as,
\begin{align}
\langle\langle\overline{C}_l\rangle\rangle_{\mathcal{S}_C}=\text{max}[0,m_l(C^l-(C_{\textrm{Th}}(l))^l)],
    \label{net_swap_mean}
\end{align}
where, $C_{\textrm{Th}}(l)$ is the length-dependent threshold concurrence and, $m_l$, is the slope when the quantity, $\langle\langle\overline{C}_l\rangle\rangle_{\mathcal{S}_C}$, is considered as a function of $C^l$. A numerical data fit yields a scaling of the threshold concurrence with the path length as,
\begin{align}
    C_{\textrm{Th}}(l)\sim (1-\xi/l), l\geq 2,
    \label{cth_mixed}
\end{align}
with, $\xi=1.35$. The slope, $m_l$, can be obtained from Eq. (\ref{cth_mixed}) by noticing that, $\langle\langle\overline{C}_l\rangle\rangle_{\mathcal{S}_C}|_{C=1}=1\forall l$, which implies, $m_l=1/(1-(C_{\textrm{Th}}(l))^l)$, resulting in values of the slope, $m_{l=2}=1.151,m_{l=3}=1.218,m_{l=4}=1.278$ and so on, increasing with $l$. The physical implication is that the average ensemble mean takes small values for input concurrences upto the threshold value, $C_{\textrm{Th}}(l)$, for a given $l$, and thereafter rapidly increases to $1$ with slopes that increase with the length, $l$. Moreover, from the expression of $C_{\text{Th}}(l)$, we see that for longer network paths, states with higher concurrence are needed along each edge to establish non-zero end-to-end entanglement. In the numerical analysis we limit ourselves to $2\leq l\leq 6$ since even large-scale quantum networks are expected to exhibit the small-world property \cite{ent_topography,q_int_brito} limiting the path lengths to this range.

Assuming that the input concurrence is above the threshold, Eq. (\ref{net_swap_mean}), implies that a quantum task with a threshold concurrence  $C_*$ \cite{wehner_qnet,zhang2022device,zapatero2023advances} may be executed using entanglement distributed along paths comprised of multi-parameter states of length up to, $l_{*}$, if
\begin{align}
m_{l_*}(C^l-(C_{\textrm{Th}}(l_*))^{l_*})\geq C_*,
\end{align}
which can be solved numerically for $l_*$ using the values for the value of thresholds given by Eq. (\ref{cth_mixed}) and the related slope.

\begin{figure}
    \centering
    \includegraphics[width=\linewidth]{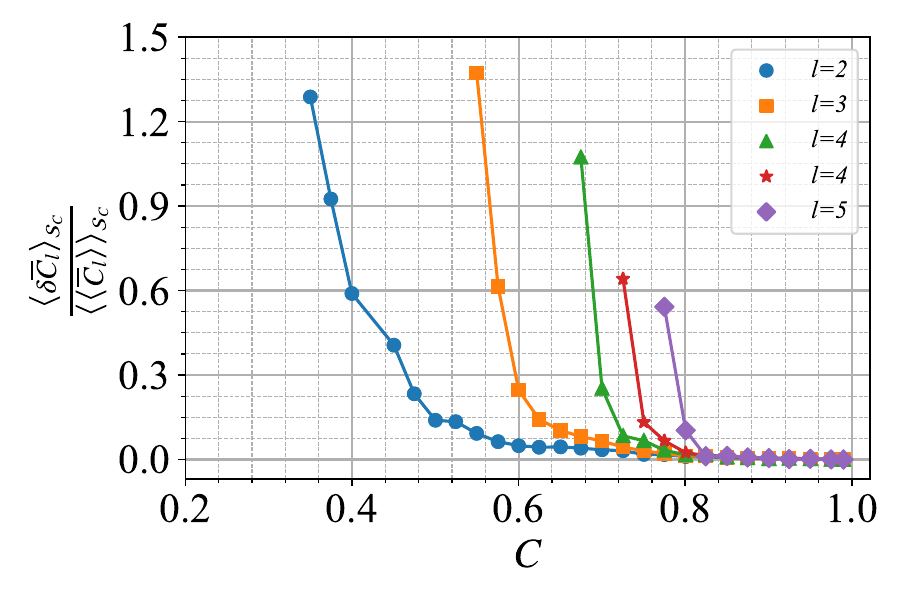}
    \caption{Ratio of the average range of the average output concurrence, $\langle\delta\overline{C}_l\rangle_{\mathcal{S}_C}$, to the average of the ensemble means, $\langle\langle\overline{C}_l\rangle\rangle_{\mathcal{S}_C}$, plotted with respect to the edge concurrence, $C$, for values above the threshold, $C> C_\textrm{Th}(l)$, for network paths comprised of multi-parameter states of the same concurrence, $C$, for different path lengths, $l=2,3,4,5,6$.}
    \label{fig:relative_range}
\end{figure}

Notice that the threshold concurrence for single-parameter states can be determined analytically for arbitrary path lengths, $l$. For pure states along the edges of a path the threshold concurrence is zero for all $l$, that is, swapping two pure states of non-zero concurrence always yields a non-zero output average concurrence. However, with Werner or Isotropic states along the edges of the path, the threshold concurrence scales with the path length as, $C_\textrm{Th}^\textrm{W}(l)=\frac{1}{2}(3^{1-1/l}-1)$, as shown in Appendix \ref{appendix_A}. Interestingly, the analytically obtained values of $C_\textrm{Th}^\textrm{W}(l)$ for Werner or Isotropic states are approximately the same as the threshold values given by Eq. (\ref{cth_mixed}) for multi-parameter states obtained by numerical analysis over the ensembles $\mathcal{E}_C(\rho)$ and $\mathcal{S}_C$. This suggests that as models for the states along the edges of a quantum network, the simpler class of Werner or Isotropic states typically used in literature \cite{ent_topography,PENstates,Santra_20191,QN_cap_trans,QN_optimal_routing} is equivalent in terms of the thresholding behavior as the more general and arguably more realistic multi-parameter network edge states.

Next, plotting in Fig.~\ref{fig:Swapping_Network}\textcolor{red}{B} the average range, $\langle\delta\overline{C}_l\rangle_{\mathcal{S}_C}$, with respect to the input concurrence, $C$, for paths of different length, $l\geq 2$, we find that the maximum value of the average range decreases with increasing path length, $l$. Further, we find that the domain of values for the input concurrence, $C\in[C_{\textrm{Th}}(l),1]$, where the average range, $\langle\delta\overline{C}_l\rangle_{\mathcal{S}_C}$, is non-zero shrinks with, $l$, since the threshold, $C_{\textrm{Th}}(l)$, moves closer to 1 as $l$ increases. Within this region the average range first increases, reaching a maximum for input concurrence close to the threshold for the given length, $C\approx C_{\textrm{Th}}(l)$, and then decreases as $C\to1$. 

One final observation from Fig. \ref{fig:Swapping_Network} is that the average range, $\langle\delta\overline{C}_l\rangle_{\mathcal{S}_C}$, may be non-zero for values of the concurrence, $C\lesssim C_{\textrm{Th}}(l)$, somewhat below the threshold concurrence, $C_{\mathrm{Th}}(l)$, though the average ensemble mean of the path concurrence, $\langle\langle\overline{C}_l\rangle\rangle_{\mathcal{S}_C}$, is close to zero near the threshold. This is consistent with the fact that there may be statistically rare instances of two states both below the threshold concurrence that yield a non-zero average output concurrence post entanglement swapping. Therefore, the threshold, $C_{\textrm{Th}}(l)$, should be interpreted as the statistically observed minimum value of concurrence for the network edge states to yield non-vanishing end-to-end entanglement, on-average.

From Fig. \ref{fig:relative_range}, we find that the magnitude of the average range relative to the average of the ensemble means, $\langle\delta\overline{C}_l\rangle_{\mathcal{S}_C}/\langle\langle\overline{C}_l\rangle\rangle_{\mathcal{S}_C}$, takes it maximum value close to the threshold concurrence, $C_{\textrm{Th}}(l)$, for a given length, $l$. Further, from the same figure we observe that this maximum value of the ratio decreases with increasing path length, $l$, with the maximum being, $~130\%$ at $C\approx0.36$ for $l=2$, and, $~50\%$ at $C\approx0.76$ for $l=6$. This implies that the output average concurrence for longer network paths becomes increasingly concentrated around the average of the ensemble mean given by Eq. (\ref{net_swap_mean}). 

The numerical study of the post-swap average output concurrence over ensembles of multi-parameter states of various purities for different path lengths, $l$, reveals that it can vary significantly over certain regions of the input concurrence, $C$ - particularly, for $C\gtrsim C_{\textrm{Th}}(l)$. Importantly, this implies that specifying the concurrences of multi-parameter states over a quantum network's edges is insufficient, by itself, for the purpose of finding optimal paths for entanglement distribution in a network - as we show in the next section.

\section{Indeterminate optimal paths over ensembles of multi-parameter states\label{section_5}}
The optimal path with regard to entanglement distribution between a pair of source (S) - destination (D) nodes in a quantum network is the path that gives the highest value of the concurrence of the end-to-end state obtained by swapping at intermediate nodes along the path. For a quantum network with multi-parameter states along its edges, the optimal path, among a set of alternate paths between a pair of S, D nodes, is indeterminate given only the concurrence values of the network edge states but not the form of their density matrices - as we show below.

\begin{figure}
    \centering
    \includegraphics[width=\linewidth]{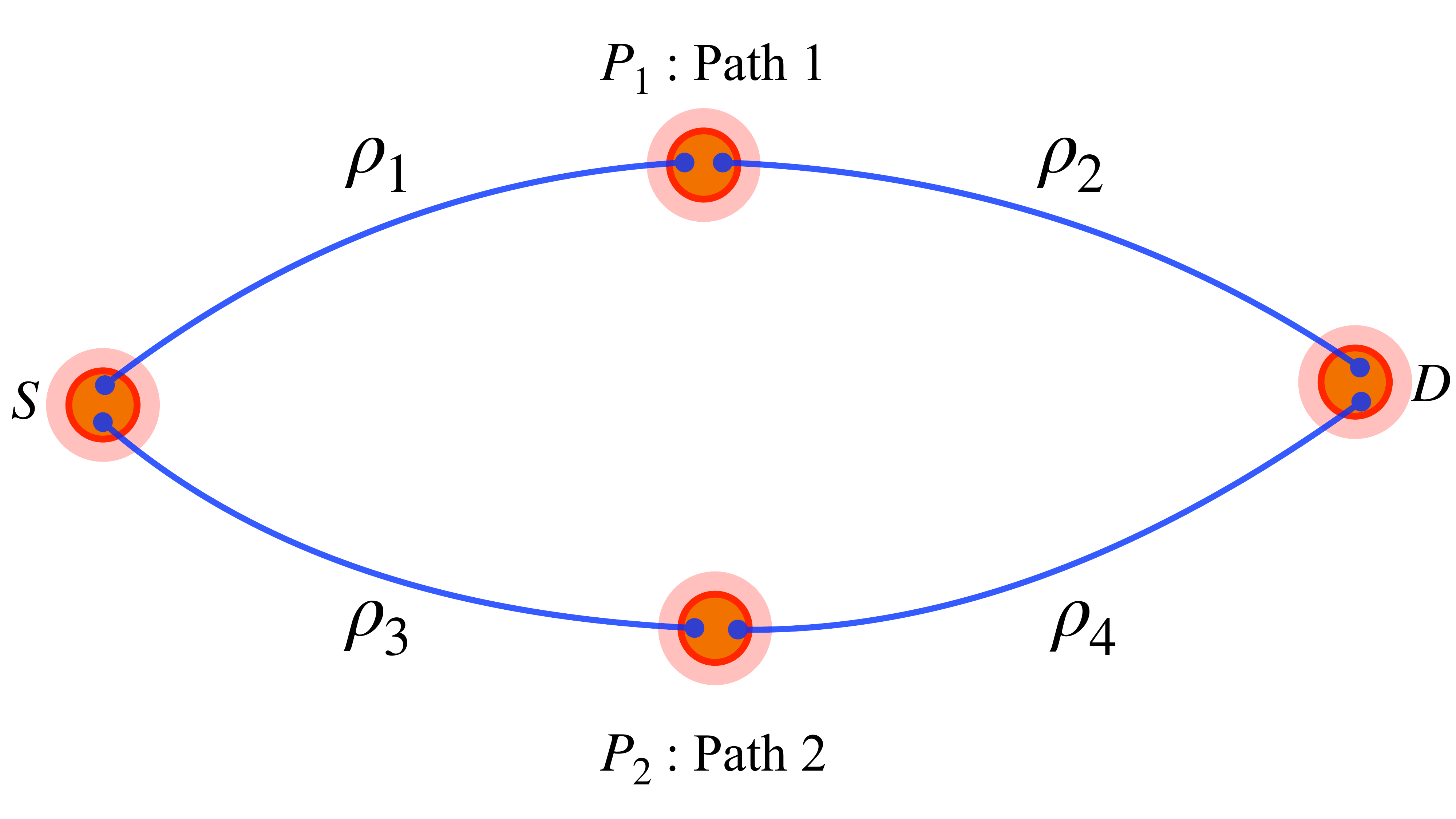}
    \caption{A pair of source-destination nodes, $S$ and $D$, in a network are connected by two alternate paths, $P_1$ and $P_2$, each of length $l = 2$, comprising two edges. All the edges are distributed with multi-parameter states. The quantum states distributed along the edges of path $P_1$ are $\rho_1$ and $\rho_2$, and those along path $P_2$ are $\rho_3$ and $\rho_4$, respectively. All four states have the same concurrence: $\mathcal{C}(\rho_1) = \mathcal{C}(\rho_2) = \mathcal{C}(\rho_3) = \mathcal{C}(\rho_4) = 0.5$.}
    \label{fig:optimal_path}
\end{figure}

Consider the situation in Fig. \ref{fig:optimal_path} where there are two alternate paths, $P_1$ and $P_2$, with two edges each, between the S-D nodes in a quantum network where the edges are characterised by multi-parameter states all with the same concurrence, $C\approx0.5$. The multi-parameter states along the two edges of the $P_1$ path are denoted by $\rho_1$ and $\rho_2$, whereas, the states along the two edges of the $P_2$ path are denoted by $\rho_3$ and $\rho_4$. We will show that the optimal path flips between, $P_1$ and $P_2$, as the edges along the paths are assigned different states picked from a subset of states within the ensemble, $\mathcal{E}_C(\rho)$. 

The fiducial state, $\rho$, that generates the ensemble is chosen to be,
 \begin{align}
\rho=\begin{pmatrix}
0.115 & -0.093 & -0.113 & -0.145\\
-0.093 & 0.373 & 0.154 & 0.320\\
-0.113 & 0.154 & 0.152 & 0.161\\
-0.145 & 0.320 & 0.161 & 0.360
\end{pmatrix},
\end{align}
with the states, $\rho_1,\rho_2,\rho_3,\rho_4$, along the edges generated by applying distinct local unitary rotations on the states along the first and second paths as follows,
\begin{align}
\rho_1(\theta_1)&=\rho_2(\theta_1)=(u_1\otimes u_1)\rho(u_1\otimes u_1)^\dagger,u_1=u(\theta_1,\theta_2=\textrm{fixed}),\nonumber\\
\rho_3(\theta_2)&=\rho_4(\theta_2)=(u_2\otimes u_2)\rho(u_2\otimes u_2)^\dagger,u_2=u(\theta_1=\textrm{fixed},\theta_2),
\label{path_states}
\end{align}
with the unitary, $u(\theta_1,\theta_2)$, given by,
\begin{align}
u(\theta_1,\theta_2)=\begin{pmatrix}
\cos(\theta_1/2) & -e^{-i\theta_2}\sin(\theta_1/2)\\
e^{i\theta_2}\sin(\theta_1/2) & \cos(\theta_1/2)
\end{pmatrix}.
\label{eq:u_mat}
\end{align}

\begin{figure}
    \centering
    \includegraphics[width=\linewidth]{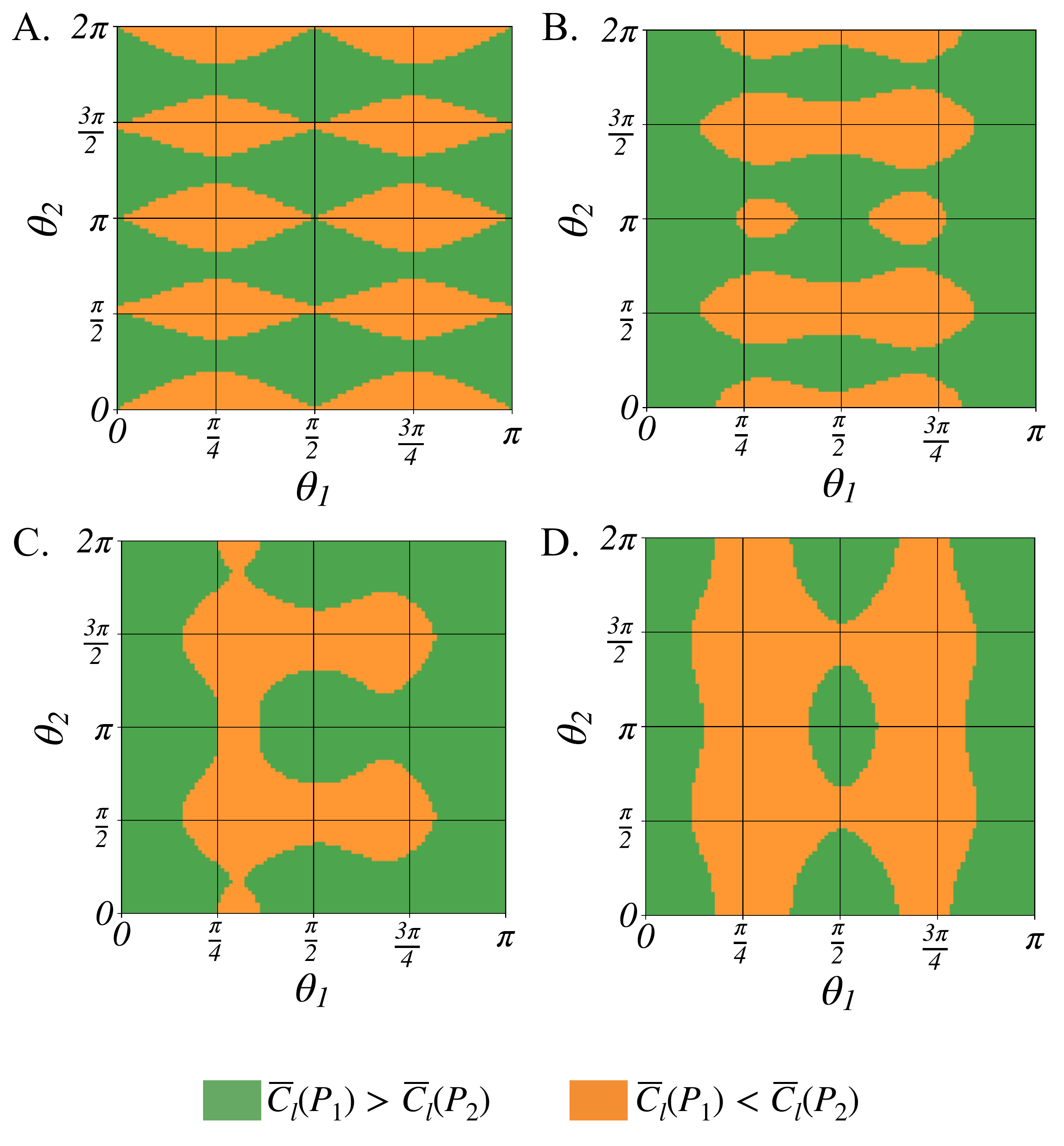}
    \caption{(Color online) Optimal choice of path for end-to-end entanglement distribution between the nodes, $S,D$, shown in Fig. \ref{fig:optimal_path} with multi-parameter states along the network edges for different values of the parameters, $\theta_1,\theta_2$, of the unitary matrix given by Eq. (\ref{eq:u_mat}). In all panels the Green (Yellow) shaded region indicates where path $P_1$ ($P_2$) yields a higher value of the average output concurrence. For all plots, the two states along the path $P_1$ are identical, $\rho_1(\theta_1)=\rho_2(\theta_1)$, as are the states along path $p_2$ with, $\rho_3(\theta_2)=\rho_4(\theta_2)$ but differ in the unitaries used to obtain these states using Eq. (\ref{path_states}).
    The unitaries used in the four panels are (A) $u_1=u(\theta_1,\theta_2=\pi/2), u_2=u(\theta_1=\pi/2,\theta_2)$ (B)  $u_1=u(\theta_1,\theta_2=\pi/3), u_2=u(\theta_1=\pi/3,\theta_2)$ (C) $u_1=u(\theta_1,\theta_2=\pi/4), u_2=u(\theta_1=\pi/4,\theta_2)$ and (D)  $u_1=u(\theta_1,\theta_2=\pi/6), u_2=u(\theta_1=\pi/6,\theta_2)$.}
    \label{fig:optimal_path_ensemble}
\end{figure}

For different choices of the states, $\rho_1(\theta_1),\rho_2(\theta_1)$ along path $P_1$, and, $\rho_3(\theta_2),\rho_4(\theta_2)$ along path $P_2$, obtained by sampling $\theta_1\in[0,\pi]$ and $\theta_2\in[0,2\pi]$ - we find that the optimal path alternates between the two choices, $P_1$ and $P_2$, as $\theta_1$ and $\theta_2$ are varied as shown in Fig. \ref{fig:optimal_path_ensemble}. This alternating behaviour of the two paths is due to the non-zero uncertainty of the average output concurrence for multi-parameter states, $\pm\langle\delta\overline{C}\rangle_{\mathcal{S}_C}/2$, when the concurrence of the network edge states is above the threshold, $C_{\textrm{Th}}(l)$, for a path of length-$l$. In contrast, when the states along the four edges are taken to be single-parameter states such as those in Subsec. \ref{section_2b}, the optimal path between S,D remains invariant over any choice of the four states from ensembles of states with fixed concurrence values. 
Remarkably, one can estimate the expected average output concurrence, $\overline{C}$, obtained over the paths, $P_1,P_2$, in case of multi-parameter network edge states using Eq. (\ref{two_swap_mean}), without needing the precise forms of the density matrices of the network edge states, while requiring only their concurrence - highlighting the utility of the numerical analysis of multi-parameter state ensembles described in Sec. (\ref{section_4}). This estimate remains uncertain upto a value of $\pm\langle\delta\overline{C}\rangle_{\mathcal{S}_C}/2$, which implies that knowledge of concurrence alone is insufficient to determine the optimality of network paths for end-to-end entanglement distribution.

\section{Discussion and Conclusions\label{section_6}}

Our analysis of entanglement swapping with multi-parameter states has revealed the scaling behaviour of the 
end-to-end entanglement of quantum network paths, $\overline{C}_l$, with network path length $l$ given only the value of the input concurrence, $C$, of the states along the edges of the path. Since for such states the average output concurrence takes a range of values, $0\leq \overline{C}_l\leq C^l$, that cannot be determined unless the density matrices of the states are provided - we resorted to a numerical analysis of ensembles of states of a given concurrence with varying purities used as inputs for the entanglement swapping operations. For such ensembles we obtained the average ensemble mean, $\langle\langle\overline{C}_l\rangle\rangle_{\mathcal{S}_C}$, and the average range, $\langle\delta\overline{C}\rangle_{\mathcal{S}_C}$, as a function of the concurrence, $C$, and the path length, $l$ - without the requirement for the precise forms of the density matrices along the edges of a quantum network. We showed that given quantum networks of multi-parameter states with only entanglement guarantees along the network edges, the estimate of the end-to-end concurrence along network paths will have uncertainties stemming from the distinct behavior under entanglement swapping of the multitude of states compatible with the given concurrence values along the edges. 

We found that there exists a path-length dependent threshold concurrence, $C_{\textrm{Th}}(l)$, for multi-parameter quantum states, such that, for $C<C_{\textrm{Th}}(l)$, the expected output concurrence vanishes, whereas, for $C\geq C_{\textrm{Th}}(l)$, it goes as $\sim m_l (C^l-(C_{\textrm{Th}}(l))^l)$ with a slope $m_l$ that increases with the path-length, $l\geq 2$. The threshold value for non-zero average output concurrence, $C_{\textrm{Th}}(l)\sim(1-\xi/l),\xi=1.35$, increases with $l$ as shown in Eq. (\ref{cth_mixed}) implying that longer paths in a quantum network require states with increasing entanglement over each edge for the expected end-to-end entanglement to be non-zero. Further, we found that the slope, $m_l$, and the threshold for a given path length, $C_{\textrm{Th}}(l)$, are related to each other as, $m_l=1/(1-(C_{\textrm{Th}}(l))^l)$, implying that for longer paths the scaling of the average output concurrence is steeper with the input concurrence.

The observed close correspondence between the threshold values, $C_{\textrm{Th}}(l)$, for multi-parameter states and those for Isotropic or Werner states, for every value of, $l$ - partially justifies the use of the latter simple, single-parameter states as the assumed network-edge states in quantum network models \cite{ent_topography,PENstates,Santra_20191,QN_cap_trans,QN_optimal_routing} - due to their similar thresholding behaviour. However, we showed that the behavior of paths with regard to their optimality for entanglement distribution showed markedly different behavior in the two cases: paths obtained by swapping multi-parameter states of a given concurrence have uncertain average output concurrence leading to indeterminate optimal paths given only entanglement guarantees along the edges of the path. This implies that entanglement routing algorithms \cite{mihir_2019,chakraborty_routing} applicable to single-parameter states, such as the pure, Isotropic or Werner state, utilising the end-to-end concurrence as the optimisation metric may not be directly suitable for obtaining optimal paths in networks with multi-parameter states along their edges. In this context, an interesting open question would be to ask: what is the minimal amount of information about the network edge states required to reliably determine the optimal paths for end-to-end entanglement distribution in a quantum network? Since, the quantum channels represented by the edges in a physical quantum network are 
likely to suffer from more general noise than the typically assumed depolarising noise that results in simple, single-parameter states such as the Isotropic or Werner state, our results point to the necessity of developing entanglement routing algorithms appropriate for quantum networks with multi-parameter network edge states.

\section{Acknowledgements} Funding from DST, Govt. of India through the SERB grant MTR/2022/000389, IITB TRUST Labs grant DO/2023-SBST002-007 and the IITB seed funding is gratefully acknowledged.

\newpage
\appendix
\section{Entanglement swapping of Werner states \label{appendix_A}}

Consider a pair of qubits $a$ and $b$ are entangled with each other and another pair of qubits $c$ and $d$ are also entangled. The two quantum states shared by qubit pairs $a-b$ and $c-d$ are Werner states described by the state parameters $\gamma_1$ and $\gamma_2$ respectively as,
\begin{align}
    \rho_{1}&=(1-\gamma_1)\ket{\Psi^-}_{ab}\bra{\Psi^-}+\frac{\gamma_1}{4}\boldsymbol{I_4},\nonumber\\
    \rho_{2}&=(1-\gamma_2)\ket{\Psi^-}_{cd}\bra{\Psi^-}+\frac{\gamma_2}{4}\boldsymbol{I_4}.
\end{align}
The concurrences of the states $\rho_{1}$ and $\rho_{2}$ can be obtained using Eq.~(\ref{werner_conc}). The probabilities of the possible BSM outcomes according to Eq.~(\ref{swapped_probability}) are, $p^{\Phi^+}=p^{\Phi^-}=p^{\Psi^+}=p^{\Psi^-}=1/4$. The corresponding possible output states after BSM according to Eq.~(\ref{swapped_state}) are given by,
\begin{align}
    \rho_\text{out}^{\Phi^\pm}&=(1-\gamma)\ket{\Phi^\pm}_{ab}\bra{\Phi^\pm}+\frac{\gamma}{4}\boldsymbol{I_4},\nonumber\\
    \rho_\text{out}^{\Psi^\pm}&=(1-\gamma)\ket{\Psi^\pm}_{ab}\bra{\Psi^\pm}+\frac{\gamma}{4}\boldsymbol{I_4},
\end{align}
where, $\gamma=\gamma_1+\gamma_2-\gamma_1\gamma_2$. The output states $\rho_\text{out}^{\Phi^\pm}$ and $\rho_\text{out}^{\Psi^\pm}$ are also Werner states and the concurrence of them are, $\mathcal{C}(\rho_\text{out}^{\Phi^\pm})=\mathcal{C}(\rho_\text{out}^{\Psi^\pm})=\text{max}(0,1-\frac{3}{2}\gamma)$. 
The output average concurrence can be obtained in terms of the input concurrences as,
\begin{align}
\overline{C}&=\text{max}\left[0,\frac{1}{3}(C_{1}+C_{2}+2C_{1}C_{2}-1)\right].
\label{ww_swap}
\end{align}
Generalizing this for swapping along a network path of length $l$, where all the edges are distributed with Werner states and the $k^\textrm{th}$ edge has concurrence, $C_k$, we obtain the average path concurrence as,
\begin{align}
    \overline{C}_l=\text{max}\left[0,\frac{3}{2}\prod_{k=1}^l\frac{1}{3}(1+2C_k)-\frac{1}{2}\right].
\end{align}

To obtain an useful network path after swapping, the average edge concurrence along the network path needs to be greater than the threshold concurrence $(C_\textrm{Th}^\textrm{W})$, which depends on the path length. In this case, where the network edges are described by Werner states, the threshold concurrence is analytically obtained as,
\begin{align}
    C_\textrm{Th}^\textrm{W}(l)=\frac{1}{2}(3^{1-1/l}-1).
    \label{cth_werner}
\end{align}
Notice from Eq.~(\ref{cth_werner}) that for $l \to \infty$, $C_\textrm{Th}^\textrm{W}\to 1$, hence for two end nodes separated by a large graph distance, the path entanglement after swapping at the intermediate nodes will be useful for quantum communication only if the edges are distributed with maximally entangled states.

\section{Entanglement swapping of pure state with mixed state \label{appendix_b}}
In Subsec. \ref{pure_mix} we have presented the results for entanglement swapping of one-parameter state with multi-parameter state. We have discussed that the output average concurrence of the post swapped state for one pure state and one X-state as inputs can be exactly obtained analytically to be the product of the two input states. From numerical investigation we show that this result is true for swapping of a pure state with any arbitrary mixed state. Here we first analytically find the average concurrence after entanglement swapping of a pure state with a general X-state. Then we present the numerical result for the swapping of a pure state with an arbitrary mixed state.

\subsection{\textbf{Swapping pure state with general X-state} \label{section_2c1}}
Consider quantum systems $a$ and $b$ along one network edge share the quantum state given by Eq.~(\ref{pure_state}) having concurrence $C_1$ and systems $c$ and $d$ along other adjacent network edge share an X-state mentioned in Eq.~(\ref{x_state}) having concurrence $C_2$.

Entanglement swapping is performed on qubits $b$ and $d$ and the probabilities of the outcomes of BSM according to Eq.~(\ref{swapped_probability}) are given by, $p^{\Phi^+}=\frac{1}{2}(\lambda(\alpha_{11}+\alpha_{22})+(1-\lambda)(\alpha_{33}+\alpha_{44}))=p^{\Phi^-}$ and $p^{\Psi^+}=\frac{1}{2}(\lambda(\alpha_{33}+\alpha_{44})+(1-\lambda)(\alpha_{11}+\alpha_{22}))=p^{\Psi^-}$. The four corresponding possible output states are,
\begin{align}
\rho_\text{out}^{\Phi^{\pm}}=\frac{1}{2p^{\Phi^\pm}}\begin{pmatrix}
        \lambda \gamma_{11} & 0 & 0 & \pm \lambda'\gamma_{14}\\
        0 & \lambda \gamma_{22} & \pm \lambda' \gamma_{23} & 0\\
        0 & \pm \lambda' \gamma_{23} & (1-\lambda) \gamma_{33} & 0\\
        \pm\lambda'\gamma_{14} & 0 & 0 & (1-\lambda) \gamma_{44}
\end{pmatrix}\nonumber\\
    \rho_\text{out}^{\Psi^{\pm}}=\frac{1}{2p^{\Psi^\pm}}\begin{pmatrix}
        \lambda \gamma_{33} & 0 & 0 & \pm \lambda' \gamma_{23}\\
        0 & \lambda \gamma_{44} & \pm \lambda' \gamma_{14} & 0\\
        0 & \pm \lambda' \gamma_{14} & (1-\lambda) \gamma_{11} & 0\\
        \pm\lambda' \gamma_{23} & 0 & 0 & (1-\lambda) \gamma_{22}
\end{pmatrix}\nonumber,
\end{align}
where, we have used $\lambda'=\sqrt{\lambda(1-\lambda)}$ for simplification. Notice that, the four possible post swapped states are also of the form of X-states. The concurrence of these states are given by, 
\begin{align}
\mathcal{C}(\rho_\text{out}^{\Phi^\pm})=&\frac{1}{p^{\Phi^+}}\text{max}[0,\lambda'\gamma_{14}-\lambda'\sqrt{\gamma_{22}\gamma_{33}},\nonumber\\
&\lambda'\gamma_{23}-\lambda'\sqrt{\gamma_{11}\gamma_{44}}]=\frac{C_{1}C_{2}}{4p^{\Phi^\pm}}\nonumber\\
\mathcal{C}(\rho_\text{out}^{\Psi^\pm})=&\frac{1}{p^{\Psi^+}}\text{max}[0,\lambda'\gamma_{23}-\lambda'\sqrt{\gamma_{11}\gamma_{44}},\nonumber\\
&\lambda'\gamma_{14}-\lambda'\sqrt{\gamma_{22}\gamma_{33}}]=\frac{C_{1}C_{2}}{4p^{\Psi^\pm}}
\end{align} 
The average concurrence of the output state according to Eq.~(\ref{avg_concurrence}) is obtained as,
\begin{align}
\overline{C}=C_{1}C_{2}.
\label{swap_px}
\end{align}
Eq.~(\ref{swap_px}) denotes that the average concurrence of the output state post entanglement swapping of a pure state with an X-state is simply the product of the concurrences of the two input states.
\subsection{\textbf{Swapping pure state with arbitrary mixed state \label{section_2c2}} \label{section_2b3}}
\begin{figure}
    \centering
    \includegraphics[width=\linewidth]{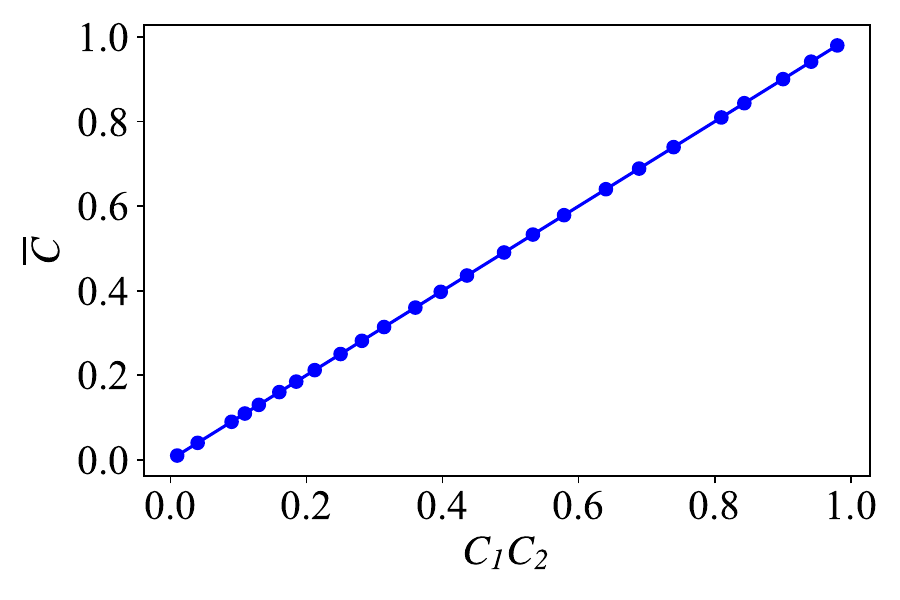}
    \caption{Average concurrence, $\overline{C}$, after entanglement swapping of $10^4$ pairs of one pure state and one arbitrary mixed state is plotted against the product, $C_{1}C_{2}$, of the input concurrences. Only a few data points are shown for clarity of the plot and the rest of the data are plotted in the solid line.}
    \label{fig:pure_mixed_swap}
\end{figure}

We have been unable to obtain an exact analytical expression for the output average concurrence after entanglement swapping of a pure state and an arbitrary mixed state. However, the numerical result presented in Fig.~\ref{fig:pure_mixed_swap} of entanglement swapping a pure state of concurrence, $C_1$, and an arbitrary mixed state of concurrence, $C_2$, suggests that the average concurrence of the post swapped output state after is the product of the two input concurrences. Hence,
\begin{equation}
    \overline{C}=C_{1}C_{2}.
    \label{pm_swap}
\end{equation}
Eq.~(\ref{pm_swap}) interestingly denote that the average concurrence of the output state can be obtained in terms of the input concurrences if any one of the input states is a pure state.

\section{Entanglement swapping of two arbitrary mixed states \label{appendix_c1}}
\subsection{Numerical method for generating the Haar random unitary ensemble \label{appendix_c1}}
To numerically generate single qubit unitary matrix, $u\in \text{U}_{\text{Haar}}(2)$, distributed according to Haar measure, we use the algorithm proposed by Mezzadri in \cite{haarmatrices_mezzadri} using QR decomposition of matrices from Ginibre ensemble. In our case of $2\times2$ unitary matrices the algorithm has the following steps :
\begin{enumerate}
    \item {\it Ginibre Matrix Generation:} The $2 \times 2$ matrices $M_1$ and $M_2$ with real numbers picked from normal distribution between 0 and 1 as their elements is generated. Combining these matrices as $M = M_1 + iM_2$, where $i=\sqrt{-1}$, yields a matrix belonging to the Ginibre ensemble.
    \item {\it QR Decomposition:} The QR decomposition of matrix $M$, yielding $M = QR$, where $Q$ is an orthogonal matrix and $R$ is an upper-triangular matrix is computed. Notably, $Q$ represents a unitary matrix, laying the foundation for generating Haar-random matrices.
    \item {\it Diagonal Matrix Generation:} A diagonal matrix $D$ by normalizing the diagonal elements of $R$, ensuring that they have unit modulus, is constructed. This step is crucial for preserving the unitarity of the resulting Haar-random matrix.
    \item {\it Haar-Random Matrix Generation:} The Haar-random matrix is, $u = Q \cdot D$.
    \item {\it Haar unitary set Generation:} Iterating step $1$ through $4$ for a large number of times, a set, $\textrm{U}_\textrm{Haar}(2)$, of Haar random unitaries is obtained.
\end{enumerate}

\subsection{\textbf{Analytical upper bound on the output average concurrence of input mixed states}}
\label{sec:ub_avg_conc}
We have discussed that the post swapped output average concurrence can not be obtained deterministically with only the knowledge of the input concurrences. We rather obtain a distribution of the output average concurrence. Here, we analytically calculate an upper bound of the output average concurrence which is not a tight upper bound.

Consider two input mixed states written in their respective optimal decompositions as,
\begin{align}
    \rho_{1}&=\sum_{i}p_{i}\ket{\xi_i}_{ab}\bra{\xi_i},\nonumber \\
    \rho_{2}&=\sum_{j}q_{j}\ket{\zeta_j}_{cd}\bra{\zeta_j}.
\end{align}
Here, $\ket{\xi_i}_{ab}=(\alpha_i\ket{00}+\beta_i\ket{01}+\gamma_i\ket{10}+\delta_i\ket{11})$ and $\ket{\zeta_j}_{cd}=(\alpha'_j\ket{00}+\beta'_j\ket{01}+\gamma'_j\ket{10}+\delta'_j\ket{11})$ are written in the computational basis. The concurrences of the two input states are, $C_{1}=2\abs{\alpha_i\delta_i-\beta_i\gamma_i} \ \text{and} \ C_{2}=2\abs{\alpha'_j\delta'_j-\beta'_j\gamma'_j}$. After measurement of qubits $b$ and $c$ in the Bell basis the probability of finding them to be in the state $\ket{\Phi+}$ according to Eq.~(\ref{swapped_probability}) is, $p^{\Phi^+}=\sum_{i,j}\frac{1}{2}p_iq_j[(\alpha_i\alpha'_j+\beta_i\gamma'_j)^2+(\alpha_i\beta'_j+\beta_i\delta'_j)^2+(\gamma_i\alpha'_j+\delta_i\gamma'_j)^2+(\gamma_i\beta'_j+\delta_i\delta'_j)^2]$
and the output state of $ad$ when $bc$ is found to be in $\ket{\Phi+}$ according to Eq.~(\ref{swapped_state}) is,
\begin{align}
    \rho^{\Phi^+}_{\textrm{out}}=\sum_{i,j}\frac{p_iq_jN^{\Phi+}_{ij}}{p^{\Phi^+}}\ket{\chi_{ij}^{\Phi^+}}_{ad}\bra{\chi^{\Phi^+}_{ij}}.
\end{align}
Here,
\begin{align}
    \ket{\chi_{ij}^{\Phi^+}}_{ad}&=\frac{\alpha_i\alpha'_j+\beta_i\gamma'_j}{\sqrt{2N^{\Phi^+}_{ij}}}\ket{00}+\frac{\alpha_i\beta'_j+\beta_i\delta'_j}{\sqrt{2N^{\Phi^+}_{ij}}}\ket{01}\nonumber\\
    &+\frac{\gamma_i\alpha'_j+\delta_i\gamma'_j}{\sqrt{2N^{\Phi^+}_{ij}}}\ket{10}+\frac{\gamma_i\beta'_j+\delta_i\delta'_j}{\sqrt{2N^{\Phi^+}_{ij}}}\ket{11},
\end{align}
and,
\begin{align}
    N^{\Phi^+}_{ij}&=\frac{1}{2}[(\alpha_i\alpha'_j+\beta_i\gamma'_j)^2+(\alpha_i\beta'_j+\beta_i\delta'_j)^2\nonumber\\
    &+(\gamma_i\alpha'_j+\delta_i\gamma'_j)^2+(\gamma_i\beta'_j+\delta_i\delta'_j)^2].
\end{align}
The concurrence of the state $\ket{\chi_{ij}^{\Phi^+}}_{ad}$ is given by,
{\small \begin{align}
    C(\ket{\chi_{ij}^{\Phi^+}}_{ad})&=2\abs{\frac{\alpha_i\alpha'_j+\beta_i\gamma'_j}{\sqrt{2N^{\Phi^+}_{ij}}}\frac{\gamma_i\beta'_j+\delta_ih_j}{\sqrt{2N^{\Phi^+}_{ij}}}-\frac{\alpha_i\beta'_j+\beta_i\delta'_j}{\sqrt{2N^{\Phi^+}_{ij}}}\frac{\gamma_i\alpha'_j+\delta_i\gamma'_j}{\sqrt{2N^{\Phi^+}_{ij}}}},\nonumber\\
    &=\frac{C_{1}C_{2}}{4N^{\Phi^+}_{ij}}.
\end{align}}
Since concurrence is a convex function of quantum states, the following inequality holds:
\begin{align}
    &C({\rho_\text{out}^{\Phi^+}})< \sum_{i,j}\frac{p_iq_jN^{\Phi^+}_{ij}}{p^{\Phi^+}}C(\ket{\chi_{ij}^{\Phi^+}}_{ad}),\nonumber\\
    \implies&C({\rho_\text{out}^{\Phi^+}})< \frac{C_{1}C_{2}}{4p^{\Phi^+}}.
\end{align}\\
Similar calculations yield : $C({\rho_\text{out}^{\Phi^-}})< \frac{C_{1}C_{2}}{4p^{\Phi^-}} \ \text{,} \ C({\rho_\text{out}^{\Psi^+}})< \frac{C_{1}C_{2}}{4p^{\Psi^+}} \ \text{and} \ C({\rho_\text{out}^{\Psi^-}})< \frac{C_{1}C_{2}}{4p^{\Psi^-}}$. Hence, the average concurrence of the output state is given by,
\begin{align}
    \overline{C} < C_{1}C_{2}.
    \label{mix_upperbound}
\end{align}
Ineq.~(\ref{mix_upperbound}) provides an upper bound on the output average concurrence obtained after swapping two mixed states which is the product of the input concurrences. It indicates that any two mixed states always swap to an average concurrence that is lower than what can be obtained by swapping with pure states.

\bibliographystyle{apsrev4-1}.
\bibliography{bibliography}

\begin{thebibliography}{47}%
\makeatletter
\providecommand \@ifxundefined [1]{%
 \@ifx{#1\undefined}
}%
\providecommand \@ifnum [1]{%
 \ifnum #1\expandafter \@firstoftwo
 \else \expandafter \@secondoftwo
 \fi
}%
\providecommand \@ifx [1]{%
 \ifx #1\expandafter \@firstoftwo
 \else \expandafter \@secondoftwo
 \fi
}%
\providecommand \natexlab [1]{#1}%
\providecommand \enquote  [1]{``#1''}%
\providecommand \bibnamefont  [1]{#1}%
\providecommand \bibfnamefont [1]{#1}%
\providecommand \citenamefont [1]{#1}%
\providecommand \href@noop [0]{\@secondoftwo}%
\providecommand \href [0]{\begingroup \@sanitize@url \@href}%
\providecommand \@href[1]{\@@startlink{#1}\@@href}%
\providecommand \@@href[1]{\endgroup#1\@@endlink}%
\providecommand \@sanitize@url [0]{\catcode `\\12\catcode `\$12\catcode
  `\&12\catcode `\#12\catcode `\^12\catcode `\_12\catcode `\%12\relax}%
\providecommand \@@startlink[1]{}%
\providecommand \@@endlink[0]{}%
\providecommand \url  [0]{\begingroup\@sanitize@url \@url }%
\providecommand \@url [1]{\endgroup\@href {#1}{\urlprefix }}%
\providecommand \urlprefix  [0]{URL }%
\providecommand \Eprint [0]{\href }%
\providecommand \doibase [0]{http://dx.doi.org/}%
\providecommand \selectlanguage [0]{\@gobble}%
\providecommand \bibinfo  [0]{\@secondoftwo}%
\providecommand \bibfield  [0]{\@secondoftwo}%
\providecommand \translation [1]{[#1]}%
\providecommand \BibitemOpen [0]{}%
\providecommand \bibitemStop [0]{}%
\providecommand \bibitemNoStop [0]{.\EOS\space}%
\providecommand \EOS [0]{\spacefactor3000\relax}%
\providecommand \BibitemShut  [1]{\csname bibitem#1\endcsname}%
\let\auto@bib@innerbib\@empty
\bibitem [{\citenamefont {\ifmmode~\dot{Z}\else \.{Z}\fi{}ukowski}\ \emph
  {et~al.}(1993)\citenamefont {\ifmmode~\dot{Z}\else \.{Z}\fi{}ukowski},
  \citenamefont {Zeilinger}, \citenamefont {Horne},\ and\ \citenamefont
  {Ekert}}]{swapping_zeillinger}%
  \BibitemOpen
  \bibfield  {author} {\bibinfo {author} {\bibfnamefont {M.}~\bibnamefont
  {\ifmmode~\dot{Z}\else \.{Z}\fi{}ukowski}}, \bibinfo {author} {\bibfnamefont
  {A.}~\bibnamefont {Zeilinger}}, \bibinfo {author} {\bibfnamefont {M.~A.}\
  \bibnamefont {Horne}}, \ and\ \bibinfo {author} {\bibfnamefont {A.~K.}\
  \bibnamefont {Ekert}},\ }\href {\doibase 10.1103/PhysRevLett.71.4287}
  {\bibfield  {journal} {\bibinfo  {journal} {Phys. Rev. Lett.}\ }\textbf
  {\bibinfo {volume} {71}},\ \bibinfo {pages} {4287} (\bibinfo {year}
  {1993})}\BibitemShut {NoStop}%
\bibitem [{\citenamefont {Bose}\ \emph {et~al.}(1998)\citenamefont {Bose},
  \citenamefont {Vedral},\ and\ \citenamefont {Knight}}]{swapping_vedral}%
  \BibitemOpen
  \bibfield  {author} {\bibinfo {author} {\bibfnamefont {S.}~\bibnamefont
  {Bose}}, \bibinfo {author} {\bibfnamefont {V.}~\bibnamefont {Vedral}}, \ and\
  \bibinfo {author} {\bibfnamefont {P.~L.}\ \bibnamefont {Knight}},\ }\href
  {\doibase 10.1103/PhysRevA.57.822} {\bibfield  {journal} {\bibinfo  {journal}
  {Phys. Rev. A}\ }\textbf {\bibinfo {volume} {57}},\ \bibinfo {pages} {822}
  (\bibinfo {year} {1998})}\BibitemShut {NoStop}%
\bibitem [{\citenamefont {Pan}\ \emph {et~al.}(1998)\citenamefont {Pan},
  \citenamefont {Bouwmeester}, \citenamefont {Weinfurter},\ and\ \citenamefont
  {Zeilinger}}]{swapping_zeillinger2}%
  \BibitemOpen
  \bibfield  {author} {\bibinfo {author} {\bibfnamefont {J.-W.}\ \bibnamefont
  {Pan}}, \bibinfo {author} {\bibfnamefont {D.}~\bibnamefont {Bouwmeester}},
  \bibinfo {author} {\bibfnamefont {H.}~\bibnamefont {Weinfurter}}, \ and\
  \bibinfo {author} {\bibfnamefont {A.}~\bibnamefont {Zeilinger}},\ }\href
  {\doibase 10.1103/PhysRevLett.80.3891} {\bibfield  {journal} {\bibinfo
  {journal} {Phys. Rev. Lett.}\ }\textbf {\bibinfo {volume} {80}},\ \bibinfo
  {pages} {3891} (\bibinfo {year} {1998})}\BibitemShut {NoStop}%
\bibitem [{\citenamefont {Kimble}(2008)}]{q_net_kimble}%
  \BibitemOpen
  \bibfield  {author} {\bibinfo {author} {\bibfnamefont {H.~J.}\ \bibnamefont
  {Kimble}},\ }\href {\doibase 10.1038/nature07127} {\bibfield  {journal}
  {\bibinfo  {journal} {Nature}\ }\textbf {\bibinfo {volume} {453}},\ \bibinfo
  {pages} {1023} (\bibinfo {year} {2008})}\BibitemShut {NoStop}%
\bibitem [{\citenamefont {Wehner}\ \emph {et~al.}(2018)\citenamefont {Wehner},
  \citenamefont {Elkouss},\ and\ \citenamefont {Hanson}}]{wehner_qnet}%
  \BibitemOpen
  \bibfield  {author} {\bibinfo {author} {\bibfnamefont {S.}~\bibnamefont
  {Wehner}}, \bibinfo {author} {\bibfnamefont {D.}~\bibnamefont {Elkouss}}, \
  and\ \bibinfo {author} {\bibfnamefont {R.}~\bibnamefont {Hanson}},\ }\href
  {\doibase 10.1126/science.aam9288} {\bibfield  {journal} {\bibinfo  {journal}
  {Science}\ }\textbf {\bibinfo {volume} {362}},\ \bibinfo {pages} {eaam9288}
  (\bibinfo {year} {2018})}\BibitemShut {NoStop}%
\bibitem [{\citenamefont {Simon}(2017)}]{simon_qnet}%
  \BibitemOpen
  \bibfield  {author} {\bibinfo {author} {\bibfnamefont {C.}~\bibnamefont
  {Simon}},\ }\href {\doibase 10.1038/s41566-017-0032-0} {\bibfield  {journal}
  {\bibinfo  {journal} {Nature Photonics}\ }\textbf {\bibinfo {volume} {11}},\
  \bibinfo {pages} {678} (\bibinfo {year} {2017})}\BibitemShut {NoStop}%
\bibitem [{\citenamefont {Briegel}\ \emph {et~al.}(1998)\citenamefont
  {Briegel}, \citenamefont {D\"ur}, \citenamefont {Cirac},\ and\ \citenamefont
  {Zoller}}]{repeater_briegel}%
  \BibitemOpen
  \bibfield  {author} {\bibinfo {author} {\bibfnamefont {H.-J.}\ \bibnamefont
  {Briegel}}, \bibinfo {author} {\bibfnamefont {W.}~\bibnamefont {D\"ur}},
  \bibinfo {author} {\bibfnamefont {J.~I.}\ \bibnamefont {Cirac}}, \ and\
  \bibinfo {author} {\bibfnamefont {P.}~\bibnamefont {Zoller}},\ }\href
  {\doibase 10.1103/PhysRevLett.81.5932} {\bibfield  {journal} {\bibinfo
  {journal} {Phys. Rev. Lett.}\ }\textbf {\bibinfo {volume} {81}},\ \bibinfo
  {pages} {5932} (\bibinfo {year} {1998})}\BibitemShut {NoStop}%
\bibitem [{\citenamefont {Muralidharan}\ \emph {et~al.}(2016)\citenamefont
  {Muralidharan}, \citenamefont {Li}, \citenamefont {Kim}, \citenamefont
  {L{\"u}tkenhaus}, \citenamefont {Lukin},\ and\ \citenamefont
  {Jiang}}]{repeater_liang}%
  \BibitemOpen
  \bibfield  {author} {\bibinfo {author} {\bibfnamefont {S.}~\bibnamefont
  {Muralidharan}}, \bibinfo {author} {\bibfnamefont {L.}~\bibnamefont {Li}},
  \bibinfo {author} {\bibfnamefont {J.}~\bibnamefont {Kim}}, \bibinfo {author}
  {\bibfnamefont {N.}~\bibnamefont {L{\"u}tkenhaus}}, \bibinfo {author}
  {\bibfnamefont {M.~D.}\ \bibnamefont {Lukin}}, \ and\ \bibinfo {author}
  {\bibfnamefont {L.}~\bibnamefont {Jiang}},\ }\href {\doibase
  10.1038/srep20463} {\bibfield  {journal} {\bibinfo  {journal} {Scientific
  Reports}\ }\textbf {\bibinfo {volume} {6}},\ \bibinfo {pages} {20463}
  (\bibinfo {year} {2016})}\BibitemShut {NoStop}%
\bibitem [{\citenamefont {Mondal}\ \emph {et~al.}(2023)\citenamefont {Mondal},
  \citenamefont {Fields}, \citenamefont {Malinovsky},\ and\ \citenamefont
  {Santra}}]{ent_topography}%
  \BibitemOpen
  \bibfield  {author} {\bibinfo {author} {\bibfnamefont {M.~S.}\ \bibnamefont
  {Mondal}}, \bibinfo {author} {\bibfnamefont {D.}~\bibnamefont {Fields}},
  \bibinfo {author} {\bibfnamefont {V.~S.}\ \bibnamefont {Malinovsky}}, \ and\
  \bibinfo {author} {\bibfnamefont {S.}~\bibnamefont {Santra}},\ }\href@noop {}
  {\enquote {\bibinfo {title} {Entanglement topography of large-scale quantum
  networks},}\ } (\bibinfo {year} {2023}),\ \Eprint
  {http://arxiv.org/abs/2312.16009} {arXiv:2312.16009 [quant-ph]} \BibitemShut
  {NoStop}%
\bibitem [{\citenamefont {Bennett}\ \emph {et~al.}(1993)\citenamefont
  {Bennett}, \citenamefont {Brassard}, \citenamefont {Cr\'epeau}, \citenamefont
  {Jozsa}, \citenamefont {Peres},\ and\ \citenamefont
  {Wootters}}]{teleportation}%
  \BibitemOpen
  \bibfield  {author} {\bibinfo {author} {\bibfnamefont {C.~H.}\ \bibnamefont
  {Bennett}}, \bibinfo {author} {\bibfnamefont {G.}~\bibnamefont {Brassard}},
  \bibinfo {author} {\bibfnamefont {C.}~\bibnamefont {Cr\'epeau}}, \bibinfo
  {author} {\bibfnamefont {R.}~\bibnamefont {Jozsa}}, \bibinfo {author}
  {\bibfnamefont {A.}~\bibnamefont {Peres}}, \ and\ \bibinfo {author}
  {\bibfnamefont {W.~K.}\ \bibnamefont {Wootters}},\ }\href {\doibase
  10.1103/PhysRevLett.70.1895} {\bibfield  {journal} {\bibinfo  {journal}
  {Phys. Rev. Lett.}\ }\textbf {\bibinfo {volume} {70}},\ \bibinfo {pages}
  {1895} (\bibinfo {year} {1993})}\BibitemShut {NoStop}%
\bibitem [{\citenamefont {Bennett}\ and\ \citenamefont
  {Brassard}(2014)}]{qkd_bennett}%
  \BibitemOpen
  \bibfield  {author} {\bibinfo {author} {\bibfnamefont {C.~H.}\ \bibnamefont
  {Bennett}}\ and\ \bibinfo {author} {\bibfnamefont {G.}~\bibnamefont
  {Brassard}},\ }\href {\doibase https://doi.org/10.1016/j.tcs.2014.05.025}
  {\bibfield  {journal} {\bibinfo  {journal} {Theoretical Computer Science}\
  }\textbf {\bibinfo {volume} {560}},\ \bibinfo {pages} {7} (\bibinfo {year}
  {2014})},\ \bibinfo {note} {theoretical Aspects of Quantum Cryptography –
  celebrating 30 years of BB84}\BibitemShut {NoStop}%
\bibitem [{\citenamefont {Ekert}(1991)}]{qkd_ekert}%
  \BibitemOpen
  \bibfield  {author} {\bibinfo {author} {\bibfnamefont {A.~K.}\ \bibnamefont
  {Ekert}},\ }\href {\doibase 10.1103/PhysRevLett.67.661} {\bibfield  {journal}
  {\bibinfo  {journal} {Phys. Rev. Lett.}\ }\textbf {\bibinfo {volume} {67}},\
  \bibinfo {pages} {661} (\bibinfo {year} {1991})}\BibitemShut {NoStop}%
\bibitem [{\citenamefont {K{\'o}m{\'a}r}\ \emph {et~al.}(2014)\citenamefont
  {K{\'o}m{\'a}r}, \citenamefont {Kessler}, \citenamefont {Bishof},
  \citenamefont {Jiang}, \citenamefont {S{\o}rensen}, \citenamefont {Ye},\ and\
  \citenamefont {Lukin}}]{dist_q_comput_1}%
  \BibitemOpen
  \bibfield  {author} {\bibinfo {author} {\bibfnamefont {P.}~\bibnamefont
  {K{\'o}m{\'a}r}}, \bibinfo {author} {\bibfnamefont {E.~M.}\ \bibnamefont
  {Kessler}}, \bibinfo {author} {\bibfnamefont {M.}~\bibnamefont {Bishof}},
  \bibinfo {author} {\bibfnamefont {L.}~\bibnamefont {Jiang}}, \bibinfo
  {author} {\bibfnamefont {A.~S.}\ \bibnamefont {S{\o}rensen}}, \bibinfo
  {author} {\bibfnamefont {J.}~\bibnamefont {Ye}}, \ and\ \bibinfo {author}
  {\bibfnamefont {M.~D.}\ \bibnamefont {Lukin}},\ }\href {\doibase
  10.1038/nphys3000} {\bibfield  {journal} {\bibinfo  {journal} {Nature
  Physics}\ }\textbf {\bibinfo {volume} {10}},\ \bibinfo {pages} {582}
  (\bibinfo {year} {2014})}\BibitemShut {NoStop}%
\bibitem [{\citenamefont {Yin}\ \emph {et~al.}(2020)\citenamefont {Yin},
  \citenamefont {Li}, \citenamefont {Liao}, \citenamefont {Yang}, \citenamefont
  {Cao}, \citenamefont {Zhang}, \citenamefont {Ren}, \citenamefont {Cai},
  \citenamefont {Liu}, \citenamefont {Li}, \citenamefont {Shu}, \citenamefont
  {Huang}, \citenamefont {Deng}, \citenamefont {Li}, \citenamefont {Zhang},
  \citenamefont {Liu}, \citenamefont {Chen}, \citenamefont {Lu}, \citenamefont
  {Wang}, \citenamefont {Xu}, \citenamefont {Wang}, \citenamefont {Peng},
  \citenamefont {Ekert},\ and\ \citenamefont {Pan}}]{dist_q_comput_2}%
  \BibitemOpen
  \bibfield  {author} {\bibinfo {author} {\bibfnamefont {J.}~\bibnamefont
  {Yin}}, \bibinfo {author} {\bibfnamefont {Y.-H.}\ \bibnamefont {Li}},
  \bibinfo {author} {\bibfnamefont {S.-K.}\ \bibnamefont {Liao}}, \bibinfo
  {author} {\bibfnamefont {M.}~\bibnamefont {Yang}}, \bibinfo {author}
  {\bibfnamefont {Y.}~\bibnamefont {Cao}}, \bibinfo {author} {\bibfnamefont
  {L.}~\bibnamefont {Zhang}}, \bibinfo {author} {\bibfnamefont {J.-G.}\
  \bibnamefont {Ren}}, \bibinfo {author} {\bibfnamefont {W.-Q.}\ \bibnamefont
  {Cai}}, \bibinfo {author} {\bibfnamefont {W.-Y.}\ \bibnamefont {Liu}},
  \bibinfo {author} {\bibfnamefont {S.-L.}\ \bibnamefont {Li}}, \bibinfo
  {author} {\bibfnamefont {R.}~\bibnamefont {Shu}}, \bibinfo {author}
  {\bibfnamefont {Y.-M.}\ \bibnamefont {Huang}}, \bibinfo {author}
  {\bibfnamefont {L.}~\bibnamefont {Deng}}, \bibinfo {author} {\bibfnamefont
  {L.}~\bibnamefont {Li}}, \bibinfo {author} {\bibfnamefont {Q.}~\bibnamefont
  {Zhang}}, \bibinfo {author} {\bibfnamefont {N.-L.}\ \bibnamefont {Liu}},
  \bibinfo {author} {\bibfnamefont {Y.-A.}\ \bibnamefont {Chen}}, \bibinfo
  {author} {\bibfnamefont {C.-Y.}\ \bibnamefont {Lu}}, \bibinfo {author}
  {\bibfnamefont {X.-B.}\ \bibnamefont {Wang}}, \bibinfo {author}
  {\bibfnamefont {F.}~\bibnamefont {Xu}}, \bibinfo {author} {\bibfnamefont
  {J.-Y.}\ \bibnamefont {Wang}}, \bibinfo {author} {\bibfnamefont {C.-Z.}\
  \bibnamefont {Peng}}, \bibinfo {author} {\bibfnamefont {A.~K.}\ \bibnamefont
  {Ekert}}, \ and\ \bibinfo {author} {\bibfnamefont {J.-W.}\ \bibnamefont
  {Pan}},\ }\href {\doibase 10.1038/s41586-020-2401-y} {\bibfield  {journal}
  {\bibinfo  {journal} {Nature}\ }\textbf {\bibinfo {volume} {582}},\ \bibinfo
  {pages} {501} (\bibinfo {year} {2020})}\BibitemShut {NoStop}%
\bibitem [{\citenamefont {Buttler}\ \emph {et~al.}(1998)\citenamefont
  {Buttler}, \citenamefont {Hughes}, \citenamefont {Kwiat}, \citenamefont
  {Lamoreaux}, \citenamefont {Luther}, \citenamefont {Morgan}, \citenamefont
  {Nordholt}, \citenamefont {Peterson},\ and\ \citenamefont
  {Simmons}}]{small_net1}%
  \BibitemOpen
  \bibfield  {author} {\bibinfo {author} {\bibfnamefont {W.~T.}\ \bibnamefont
  {Buttler}}, \bibinfo {author} {\bibfnamefont {R.~J.}\ \bibnamefont {Hughes}},
  \bibinfo {author} {\bibfnamefont {P.~G.}\ \bibnamefont {Kwiat}}, \bibinfo
  {author} {\bibfnamefont {S.~K.}\ \bibnamefont {Lamoreaux}}, \bibinfo {author}
  {\bibfnamefont {G.~G.}\ \bibnamefont {Luther}}, \bibinfo {author}
  {\bibfnamefont {G.~L.}\ \bibnamefont {Morgan}}, \bibinfo {author}
  {\bibfnamefont {J.~E.}\ \bibnamefont {Nordholt}}, \bibinfo {author}
  {\bibfnamefont {C.~G.}\ \bibnamefont {Peterson}}, \ and\ \bibinfo {author}
  {\bibfnamefont {C.~M.}\ \bibnamefont {Simmons}},\ }\href {\doibase
  10.1103/PhysRevLett.81.3283} {\bibfield  {journal} {\bibinfo  {journal}
  {Phys. Rev. Lett.}\ }\textbf {\bibinfo {volume} {81}},\ \bibinfo {pages}
  {3283} (\bibinfo {year} {1998})}\BibitemShut {NoStop}%
\bibitem [{\citenamefont {Jouguet}\ \emph {et~al.}(2013)\citenamefont
  {Jouguet}, \citenamefont {Kunz-Jacques}, \citenamefont {Leverrier},
  \citenamefont {Grangier},\ and\ \citenamefont {Diamanti}}]{small_net2}%
  \BibitemOpen
  \bibfield  {author} {\bibinfo {author} {\bibfnamefont {P.}~\bibnamefont
  {Jouguet}}, \bibinfo {author} {\bibfnamefont {S.}~\bibnamefont
  {Kunz-Jacques}}, \bibinfo {author} {\bibfnamefont {A.}~\bibnamefont
  {Leverrier}}, \bibinfo {author} {\bibfnamefont {P.}~\bibnamefont {Grangier}},
  \ and\ \bibinfo {author} {\bibfnamefont {E.}~\bibnamefont {Diamanti}},\
  }\href {\doibase 10.1038/nphoton.2013.63} {\bibfield  {journal} {\bibinfo
  {journal} {Nature Photonics}\ }\textbf {\bibinfo {volume} {7}},\ \bibinfo
  {pages} {378} (\bibinfo {year} {2013})}\BibitemShut {NoStop}%
\bibitem [{\citenamefont {et~al.}(2009)}]{small_net3}%
  \BibitemOpen
  \bibfield  {author} {\bibinfo {author} {\bibfnamefont {M.~P.}\ \bibnamefont
  {et~al.}},\ }\href {\doibase 10.1088/1367-2630/11/7/075001} {\bibfield
  {journal} {\bibinfo  {journal} {New Journal of Physics}\ }\textbf {\bibinfo
  {volume} {11}},\ \bibinfo {pages} {075001} (\bibinfo {year}
  {2009})}\BibitemShut {NoStop}%
\bibitem [{\citenamefont {et~al.}(2011{\natexlab{a}})}]{small_net4}%
  \BibitemOpen
  \bibfield  {author} {\bibinfo {author} {\bibfnamefont {M.~S.}\ \bibnamefont
  {et~al.}},\ }\href {\doibase 10.1364/OE.19.010387} {\bibfield  {journal}
  {\bibinfo  {journal} {Opt. Express}\ }\textbf {\bibinfo {volume} {19}},\
  \bibinfo {pages} {10387} (\bibinfo {year} {2011}{\natexlab{a}})}\BibitemShut
  {NoStop}%
\bibitem [{\citenamefont {et~al.}(2011{\natexlab{b}})}]{small_net5}%
  \BibitemOpen
  \bibfield  {author} {\bibinfo {author} {\bibfnamefont {D.~S.}\ \bibnamefont
  {et~al.}},\ }\href {\doibase 10.1088/1367-2630/13/12/123001} {\bibfield
  {journal} {\bibinfo  {journal} {New Journal of Physics}\ }\textbf {\bibinfo
  {volume} {13}},\ \bibinfo {pages} {123001} (\bibinfo {year}
  {2011}{\natexlab{b}})}\BibitemShut {NoStop}%
\bibitem [{\citenamefont {et~al.}(2014)}]{small_net6}%
  \BibitemOpen
  \bibfield  {author} {\bibinfo {author} {\bibfnamefont {S.~W.}\ \bibnamefont
  {et~al.}},\ }\href {\doibase 10.1364/OE.22.021739} {\bibfield  {journal}
  {\bibinfo  {journal} {Opt. Express}\ }\textbf {\bibinfo {volume} {22}},\
  \bibinfo {pages} {21739} (\bibinfo {year} {2014})}\BibitemShut {NoStop}%
\bibitem [{\citenamefont {et~al.}(2017)}]{satelite_ent_dist}%
  \BibitemOpen
  \bibfield  {author} {\bibinfo {author} {\bibfnamefont {J.~Y.}\ \bibnamefont
  {et~al.}},\ }\href {\doibase 10.1126/science.aan3211} {\bibfield  {journal}
  {\bibinfo  {journal} {Science}\ }\textbf {\bibinfo {volume} {356}},\ \bibinfo
  {pages} {1140} (\bibinfo {year} {2017})},\ \Eprint
  {http://arxiv.org/abs/https://www.science.org/doi/pdf/10.1126/science.aan3211}
  {https://www.science.org/doi/pdf/10.1126/science.aan3211} \BibitemShut
  {NoStop}%
\bibitem [{\citenamefont {Zhang}\ \emph {et~al.}(2022)\citenamefont {Zhang},
  \citenamefont {van Leent}, \citenamefont {Redeker}, \citenamefont {Garthoff},
  \citenamefont {Schwonnek}, \citenamefont {Fertig}, \citenamefont {Eppelt},
  \citenamefont {Rosenfeld}, \citenamefont {Scarani}, \citenamefont {Lim} \emph
  {et~al.}}]{zhang2022device}%
  \BibitemOpen
  \bibfield  {author} {\bibinfo {author} {\bibfnamefont {W.}~\bibnamefont
  {Zhang}}, \bibinfo {author} {\bibfnamefont {T.}~\bibnamefont {van Leent}},
  \bibinfo {author} {\bibfnamefont {K.}~\bibnamefont {Redeker}}, \bibinfo
  {author} {\bibfnamefont {R.}~\bibnamefont {Garthoff}}, \bibinfo {author}
  {\bibfnamefont {R.}~\bibnamefont {Schwonnek}}, \bibinfo {author}
  {\bibfnamefont {F.}~\bibnamefont {Fertig}}, \bibinfo {author} {\bibfnamefont
  {S.}~\bibnamefont {Eppelt}}, \bibinfo {author} {\bibfnamefont
  {W.}~\bibnamefont {Rosenfeld}}, \bibinfo {author} {\bibfnamefont
  {V.}~\bibnamefont {Scarani}}, \bibinfo {author} {\bibfnamefont {C.~C.-W.}\
  \bibnamefont {Lim}},  \emph {et~al.},\ }\href
  {https://www.nature.com/articles/s41586-022-04891-y} {\bibfield  {journal}
  {\bibinfo  {journal} {Nature}\ }\textbf {\bibinfo {volume} {607}},\ \bibinfo
  {pages} {687} (\bibinfo {year} {2022})}\BibitemShut {NoStop}%
\bibitem [{\citenamefont {Zapatero}\ \emph {et~al.}(2023)\citenamefont
  {Zapatero}, \citenamefont {van Leent}, \citenamefont {Arnon-Friedman},
  \citenamefont {Liu}, \citenamefont {Zhang}, \citenamefont {Weinfurter},\ and\
  \citenamefont {Curty}}]{zapatero2023advances}%
  \BibitemOpen
  \bibfield  {author} {\bibinfo {author} {\bibfnamefont {V.}~\bibnamefont
  {Zapatero}}, \bibinfo {author} {\bibfnamefont {T.}~\bibnamefont {van Leent}},
  \bibinfo {author} {\bibfnamefont {R.}~\bibnamefont {Arnon-Friedman}},
  \bibinfo {author} {\bibfnamefont {W.-Z.}\ \bibnamefont {Liu}}, \bibinfo
  {author} {\bibfnamefont {Q.}~\bibnamefont {Zhang}}, \bibinfo {author}
  {\bibfnamefont {H.}~\bibnamefont {Weinfurter}}, \ and\ \bibinfo {author}
  {\bibfnamefont {M.}~\bibnamefont {Curty}},\ }\href
  {https://www.nature.com/articles/s41534-023-00684-x} {\bibfield  {journal}
  {\bibinfo  {journal} {npj quantum information}\ }\textbf {\bibinfo {volume}
  {9}},\ \bibinfo {pages} {10} (\bibinfo {year} {2023})}\BibitemShut {NoStop}%
\bibitem [{\citenamefont {Leone}\ \emph {et~al.}(2024)\citenamefont {Leone},
  \citenamefont {Miller}, \citenamefont {Singh}, \citenamefont {Langford},\
  and\ \citenamefont {Rohde}}]{leone2024costvectoranalysis}%
  \BibitemOpen
  \bibfield  {author} {\bibinfo {author} {\bibfnamefont {H.}~\bibnamefont
  {Leone}}, \bibinfo {author} {\bibfnamefont {N.~R.}\ \bibnamefont {Miller}},
  \bibinfo {author} {\bibfnamefont {D.}~\bibnamefont {Singh}}, \bibinfo
  {author} {\bibfnamefont {N.~K.}\ \bibnamefont {Langford}}, \ and\ \bibinfo
  {author} {\bibfnamefont {P.~P.}\ \bibnamefont {Rohde}},\ }\href
  {https://arxiv.org/abs/2105.00418} {\enquote {\bibinfo {title} {Cost vector
  analysis \& multi-path entanglement routing in quantum networks},}\ }
  (\bibinfo {year} {2024}),\ \Eprint {http://arxiv.org/abs/2105.00418}
  {arXiv:2105.00418 [quant-ph]} \BibitemShut {NoStop}%
\bibitem [{\citenamefont {Mondal}\ and\ \citenamefont
  {Santra}(2024)}]{mep_conf_sohel}%
  \BibitemOpen
  \bibfield  {author} {\bibinfo {author} {\bibfnamefont {M.~S.}\ \bibnamefont
  {Mondal}}\ and\ \bibinfo {author} {\bibfnamefont {S.}~\bibnamefont
  {Santra}},\ }in\ \href {\doibase 10.1109/QCNC62729.2024.00055} {\emph
  {\bibinfo {booktitle} {2024 International Conference on Quantum
  Communications, Networking, and Computing (QCNC)}}}\ (\bibinfo {year}
  {2024})\ pp.\ \bibinfo {pages} {312--319}\BibitemShut {NoStop}%
\bibitem [{\citenamefont {Bala}\ \emph {et~al.}(2025)\citenamefont {Bala},
  \citenamefont {Mondal},\ and\ \citenamefont {Santra}}]{mep_sohel}%
  \BibitemOpen
  \bibfield  {author} {\bibinfo {author} {\bibfnamefont {R.}~\bibnamefont
  {Bala}}, \bibinfo {author} {\bibfnamefont {M.~S.}\ \bibnamefont {Mondal}}, \
  and\ \bibinfo {author} {\bibfnamefont {S.}~\bibnamefont {Santra}},\ }\href
  {https://arxiv.org/abs/2502.09011} {\enquote {\bibinfo {title} {Statistical
  analysis of multipath entanglement purification in quantum networks},}\ }
  (\bibinfo {year} {2025}),\ \Eprint {http://arxiv.org/abs/2502.09011}
  {arXiv:2502.09011 [quant-ph]} \BibitemShut {NoStop}%
\bibitem [{\citenamefont {Bergou}\ \emph {et~al.}(2021)\citenamefont {Bergou},
  \citenamefont {Fields}, \citenamefont {Hillery}, \citenamefont {Santra},\
  and\ \citenamefont {Malinovsky}}]{avgconcurrence_santra}%
  \BibitemOpen
  \bibfield  {author} {\bibinfo {author} {\bibfnamefont {J.~A.}\ \bibnamefont
  {Bergou}}, \bibinfo {author} {\bibfnamefont {D.}~\bibnamefont {Fields}},
  \bibinfo {author} {\bibfnamefont {M.}~\bibnamefont {Hillery}}, \bibinfo
  {author} {\bibfnamefont {S.}~\bibnamefont {Santra}}, \ and\ \bibinfo {author}
  {\bibfnamefont {V.~S.}\ \bibnamefont {Malinovsky}},\ }\href {\doibase
  10.1103/PhysRevA.104.022425} {\bibfield  {journal} {\bibinfo  {journal}
  {Phys. Rev. A}\ }\textbf {\bibinfo {volume} {104}},\ \bibinfo {pages}
  {022425} (\bibinfo {year} {2021})}\BibitemShut {NoStop}%
\bibitem [{\citenamefont {Orieux}\ \emph {et~al.}(2013)\citenamefont {Orieux},
  \citenamefont {Sansoni}, \citenamefont {Persechino}, \citenamefont
  {Mataloni}, \citenamefont {Rossi},\ and\ \citenamefont
  {Macchiavello}}]{exp_channel_charac}%
  \BibitemOpen
  \bibfield  {author} {\bibinfo {author} {\bibfnamefont {A.}~\bibnamefont
  {Orieux}}, \bibinfo {author} {\bibfnamefont {L.}~\bibnamefont {Sansoni}},
  \bibinfo {author} {\bibfnamefont {M.}~\bibnamefont {Persechino}}, \bibinfo
  {author} {\bibfnamefont {P.}~\bibnamefont {Mataloni}}, \bibinfo {author}
  {\bibfnamefont {M.}~\bibnamefont {Rossi}}, \ and\ \bibinfo {author}
  {\bibfnamefont {C.}~\bibnamefont {Macchiavello}},\ }\href {\doibase
  10.1103/PhysRevLett.111.220501} {\bibfield  {journal} {\bibinfo  {journal}
  {Phys. Rev. Lett.}\ }\textbf {\bibinfo {volume} {111}},\ \bibinfo {pages}
  {220501} (\bibinfo {year} {2013})}\BibitemShut {NoStop}%
\bibitem [{\citenamefont {Pears~Stefano}\ \emph {et~al.}(2021)\citenamefont
  {Pears~Stefano}, \citenamefont {Perito}, \citenamefont {Varga}, \citenamefont
  {Reb\'on},\ and\ \citenamefont {Iemmi}}]{exp_charac_process}%
  \BibitemOpen
  \bibfield  {author} {\bibinfo {author} {\bibfnamefont {Q.}~\bibnamefont
  {Pears~Stefano}}, \bibinfo {author} {\bibfnamefont {I.}~\bibnamefont
  {Perito}}, \bibinfo {author} {\bibfnamefont {J.~J.~M.}\ \bibnamefont
  {Varga}}, \bibinfo {author} {\bibfnamefont {L.}~\bibnamefont {Reb\'on}}, \
  and\ \bibinfo {author} {\bibfnamefont {C.}~\bibnamefont {Iemmi}},\ }\href
  {\doibase 10.1103/PhysRevA.103.052438} {\bibfield  {journal} {\bibinfo
  {journal} {Phys. Rev. A}\ }\textbf {\bibinfo {volume} {103}},\ \bibinfo
  {pages} {052438} (\bibinfo {year} {2021})}\BibitemShut {NoStop}%
\bibitem [{\citenamefont {Kucera}\ \emph {et~al.}(2024)\citenamefont {Kucera},
  \citenamefont {Haen}, \citenamefont {Arensk{\"o}tter}, \citenamefont {Bauer},
  \citenamefont {Meiers}, \citenamefont {Sch{\"a}fer}, \citenamefont {Boland},
  \citenamefont {Yahyapour}, \citenamefont {Lessing}, \citenamefont
  {Holzwarth}, \citenamefont {Becher},\ and\ \citenamefont
  {Eschner}}]{fiber_qkd_time_dep}%
  \BibitemOpen
  \bibfield  {author} {\bibinfo {author} {\bibfnamefont {S.}~\bibnamefont
  {Kucera}}, \bibinfo {author} {\bibfnamefont {C.}~\bibnamefont {Haen}},
  \bibinfo {author} {\bibfnamefont {E.}~\bibnamefont {Arensk{\"o}tter}},
  \bibinfo {author} {\bibfnamefont {T.}~\bibnamefont {Bauer}}, \bibinfo
  {author} {\bibfnamefont {J.}~\bibnamefont {Meiers}}, \bibinfo {author}
  {\bibfnamefont {M.}~\bibnamefont {Sch{\"a}fer}}, \bibinfo {author}
  {\bibfnamefont {R.}~\bibnamefont {Boland}}, \bibinfo {author} {\bibfnamefont
  {M.}~\bibnamefont {Yahyapour}}, \bibinfo {author} {\bibfnamefont
  {M.}~\bibnamefont {Lessing}}, \bibinfo {author} {\bibfnamefont
  {R.}~\bibnamefont {Holzwarth}}, \bibinfo {author} {\bibfnamefont
  {C.}~\bibnamefont {Becher}}, \ and\ \bibinfo {author} {\bibfnamefont
  {J.}~\bibnamefont {Eschner}},\ }\href {\doibase 10.1038/s41534-024-00886-x}
  {\bibfield  {journal} {\bibinfo  {journal} {npj Quantum Information}\
  }\textbf {\bibinfo {volume} {10}},\ \bibinfo {pages} {88} (\bibinfo {year}
  {2024})}\BibitemShut {NoStop}%
\bibitem [{\citenamefont {Karakosta-Amarantidou}\ \emph
  {et~al.}(2025)\citenamefont {Karakosta-Amarantidou}, \citenamefont {Yehia},\
  and\ \citenamefont {Schiavon}}]{free_space_time_dep}%
  \BibitemOpen
  \bibfield  {author} {\bibinfo {author} {\bibfnamefont {I.}~\bibnamefont
  {Karakosta-Amarantidou}}, \bibinfo {author} {\bibfnamefont {R.}~\bibnamefont
  {Yehia}}, \ and\ \bibinfo {author} {\bibfnamefont {M.}~\bibnamefont
  {Schiavon}},\ }\href {\doibase 10.1103/PhysRevResearch.7.023199} {\bibfield
  {journal} {\bibinfo  {journal} {Phys. Rev. Res.}\ }\textbf {\bibinfo {volume}
  {7}},\ \bibinfo {pages} {023199} (\bibinfo {year} {2025})}\BibitemShut
  {NoStop}%
\bibitem [{\citenamefont {Riccardi}\ \emph {et~al.}(2021)\citenamefont
  {Riccardi}, \citenamefont {Antonelli}, \citenamefont {Jones},\ and\
  \citenamefont {Brodsky}}]{brodsky_channel}%
  \BibitemOpen
  \bibfield  {author} {\bibinfo {author} {\bibfnamefont {G.}~\bibnamefont
  {Riccardi}}, \bibinfo {author} {\bibfnamefont {C.}~\bibnamefont {Antonelli}},
  \bibinfo {author} {\bibfnamefont {D.~E.}\ \bibnamefont {Jones}}, \ and\
  \bibinfo {author} {\bibfnamefont {M.}~\bibnamefont {Brodsky}},\ }\href
  {\doibase 10.1103/PhysRevApplied.15.014060} {\bibfield  {journal} {\bibinfo
  {journal} {Phys. Rev. Appl.}\ }\textbf {\bibinfo {volume} {15}},\ \bibinfo
  {pages} {014060} (\bibinfo {year} {2021})}\BibitemShut {NoStop}%
\bibitem [{\citenamefont {Kirby}\ \emph {et~al.}(2016)\citenamefont {Kirby},
  \citenamefont {Santra}, \citenamefont {Malinovsky},\ and\ \citenamefont
  {Brodsky}}]{swapping_santra}%
  \BibitemOpen
  \bibfield  {author} {\bibinfo {author} {\bibfnamefont {B.~T.}\ \bibnamefont
  {Kirby}}, \bibinfo {author} {\bibfnamefont {S.}~\bibnamefont {Santra}},
  \bibinfo {author} {\bibfnamefont {V.~S.}\ \bibnamefont {Malinovsky}}, \ and\
  \bibinfo {author} {\bibfnamefont {M.}~\bibnamefont {Brodsky}},\ }\href
  {\doibase 10.1103/PhysRevA.94.012336} {\bibfield  {journal} {\bibinfo
  {journal} {Phys. Rev. A}\ }\textbf {\bibinfo {volume} {94}},\ \bibinfo
  {pages} {012336} (\bibinfo {year} {2016})}\BibitemShut {NoStop}%
\bibitem [{\citenamefont {Hill}\ and\ \citenamefont
  {Wootters}(1997)}]{concurrence_wooters1}%
  \BibitemOpen
  \bibfield  {author} {\bibinfo {author} {\bibfnamefont {S.~A.}\ \bibnamefont
  {Hill}}\ and\ \bibinfo {author} {\bibfnamefont {W.~K.}\ \bibnamefont
  {Wootters}},\ }\href {\doibase 10.1103/PhysRevLett.78.5022} {\bibfield
  {journal} {\bibinfo  {journal} {Phys. Rev. Lett.}\ }\textbf {\bibinfo
  {volume} {78}},\ \bibinfo {pages} {5022} (\bibinfo {year}
  {1997})}\BibitemShut {NoStop}%
\bibitem [{\citenamefont {Wootters}(1998)}]{concurrence_wooters2}%
  \BibitemOpen
  \bibfield  {author} {\bibinfo {author} {\bibfnamefont {W.~K.}\ \bibnamefont
  {Wootters}},\ }\href {\doibase 10.1103/PhysRevLett.80.2245} {\bibfield
  {journal} {\bibinfo  {journal} {Phys. Rev. Lett.}\ }\textbf {\bibinfo
  {volume} {80}},\ \bibinfo {pages} {2245} (\bibinfo {year}
  {1998})}\BibitemShut {NoStop}%
\bibitem [{\citenamefont {Nielsen}\ and\ \citenamefont
  {Chuang}(2010)}]{nielsen_book}%
  \BibitemOpen
  \bibfield  {author} {\bibinfo {author} {\bibfnamefont {M.~A.}\ \bibnamefont
  {Nielsen}}\ and\ \bibinfo {author} {\bibfnamefont {I.~L.}\ \bibnamefont
  {Chuang}},\ }\href@noop {} {\emph {\bibinfo {title} {Quantum Computation and
  Quantum Information}}},\ \bibinfo {edition} {10th}\ ed.\ (\bibinfo
  {publisher} {Cambridge University Press},\ \bibinfo {year}
  {2010})\BibitemShut {NoStop}%
\bibitem [{\citenamefont {Werner}(1989)}]{werner_state}%
  \BibitemOpen
  \bibfield  {author} {\bibinfo {author} {\bibfnamefont {R.~F.}\ \bibnamefont
  {Werner}},\ }\href {\doibase 10.1103/PhysRevA.40.4277} {\bibfield  {journal}
  {\bibinfo  {journal} {Phys. Rev. A}\ }\textbf {\bibinfo {volume} {40}},\
  \bibinfo {pages} {4277} (\bibinfo {year} {1989})}\BibitemShut {NoStop}%
\bibitem [{\citenamefont {Terhal}\ and\ \citenamefont
  {Vollbrecht}(2000)}]{isotropic_state}%
  \BibitemOpen
  \bibfield  {author} {\bibinfo {author} {\bibfnamefont {B.~M.}\ \bibnamefont
  {Terhal}}\ and\ \bibinfo {author} {\bibfnamefont {K.~G.~H.}\ \bibnamefont
  {Vollbrecht}},\ }\href {\doibase 10.1103/PhysRevLett.85.2625} {\bibfield
  {journal} {\bibinfo  {journal} {Phys. Rev. Lett.}\ }\textbf {\bibinfo
  {volume} {85}},\ \bibinfo {pages} {2625} (\bibinfo {year}
  {2000})}\BibitemShut {NoStop}%
\bibitem [{\citenamefont {Cong}\ and\ \citenamefont
  {Xu}(2025)}]{swapping_cong}%
  \BibitemOpen
  \bibfield  {author} {\bibinfo {author} {\bibfnamefont {R.-Z.}\ \bibnamefont
  {Cong}}\ and\ \bibinfo {author} {\bibfnamefont {S.}~\bibnamefont {Xu}},\
  }\href {\doibase 10.1103/PhysRevA.111.012409} {\bibfield  {journal} {\bibinfo
   {journal} {Phys. Rev. A}\ }\textbf {\bibinfo {volume} {111}},\ \bibinfo
  {pages} {012409} (\bibinfo {year} {2025})}\BibitemShut {NoStop}%
\bibitem [{\citenamefont {Brito}\ \emph {et~al.}(2020)\citenamefont {Brito},
  \citenamefont {Canabarro}, \citenamefont {Chaves},\ and\ \citenamefont
  {Cavalcanti}}]{q_int_brito}%
  \BibitemOpen
  \bibfield  {author} {\bibinfo {author} {\bibfnamefont {S.}~\bibnamefont
  {Brito}}, \bibinfo {author} {\bibfnamefont {A.}~\bibnamefont {Canabarro}},
  \bibinfo {author} {\bibfnamefont {R.}~\bibnamefont {Chaves}}, \ and\ \bibinfo
  {author} {\bibfnamefont {D.}~\bibnamefont {Cavalcanti}},\ }\href {\doibase
  10.1103/PhysRevLett.124.210501} {\bibfield  {journal} {\bibinfo  {journal}
  {Phys. Rev. Lett.}\ }\textbf {\bibinfo {volume} {124}},\ \bibinfo {pages}
  {210501} (\bibinfo {year} {2020})}\BibitemShut {NoStop}%
\bibitem [{\citenamefont {Contreras-Tejada}\ \emph {et~al.}(2022)\citenamefont
  {Contreras-Tejada}, \citenamefont {Palazuelos},\ and\ \citenamefont
  {de~Vicente}}]{PENstates}%
  \BibitemOpen
  \bibfield  {author} {\bibinfo {author} {\bibfnamefont {P.}~\bibnamefont
  {Contreras-Tejada}}, \bibinfo {author} {\bibfnamefont {C.}~\bibnamefont
  {Palazuelos}}, \ and\ \bibinfo {author} {\bibfnamefont {J.~I.}\ \bibnamefont
  {de~Vicente}},\ }\href {\doibase 10.1103/PhysRevLett.128.220501} {\bibfield
  {journal} {\bibinfo  {journal} {Phys. Rev. Lett.}\ }\textbf {\bibinfo
  {volume} {128}},\ \bibinfo {pages} {220501} (\bibinfo {year}
  {2022})}\BibitemShut {NoStop}%
\bibitem [{\citenamefont {Santra}\ \emph {et~al.}(2019)\citenamefont {Santra},
  \citenamefont {Jiang},\ and\ \citenamefont {Malinovsky}}]{Santra_20191}%
  \BibitemOpen
  \bibfield  {author} {\bibinfo {author} {\bibfnamefont {S.}~\bibnamefont
  {Santra}}, \bibinfo {author} {\bibfnamefont {L.}~\bibnamefont {Jiang}}, \
  and\ \bibinfo {author} {\bibfnamefont {V.~S.}\ \bibnamefont {Malinovsky}},\
  }\href {\doibase 10.1088/2058-9565/ab0bc2} {\bibfield  {journal} {\bibinfo
  {journal} {Quantum Science and Technology}\ }\textbf {\bibinfo {volume}
  {4}},\ \bibinfo {pages} {025010} (\bibinfo {year} {2019})}\BibitemShut
  {NoStop}%
\bibitem [{\citenamefont {Zhuang}\ and\ \citenamefont
  {Zhang}(2021)}]{QN_cap_trans}%
  \BibitemOpen
  \bibfield  {author} {\bibinfo {author} {\bibfnamefont {Q.}~\bibnamefont
  {Zhuang}}\ and\ \bibinfo {author} {\bibfnamefont {B.}~\bibnamefont {Zhang}},\
  }\href {\doibase 10.1103/PhysRevA.104.022608} {\bibfield  {journal} {\bibinfo
   {journal} {Phys. Rev. A}\ }\textbf {\bibinfo {volume} {104}},\ \bibinfo
  {pages} {022608} (\bibinfo {year} {2021})}\BibitemShut {NoStop}%
\bibitem [{\citenamefont {Caleffi}(2017)}]{QN_optimal_routing}%
  \BibitemOpen
  \bibfield  {author} {\bibinfo {author} {\bibfnamefont {M.}~\bibnamefont
  {Caleffi}},\ }\href {\doibase 10.1109/ACCESS.2017.2763325} {\bibfield
  {journal} {\bibinfo  {journal} {IEEE Access}\ }\textbf {\bibinfo {volume}
  {5}},\ \bibinfo {pages} {22299} (\bibinfo {year} {2017})}\BibitemShut
  {NoStop}%
\bibitem [{\citenamefont {Pant}\ \emph {et~al.}(2019)\citenamefont {Pant},
  \citenamefont {Krovi}, \citenamefont {Towsley}, \citenamefont {Tassiulas},
  \citenamefont {Jiang}, \citenamefont {Basu}, \citenamefont {Englund},\ and\
  \citenamefont {Guha}}]{mihir_2019}%
  \BibitemOpen
  \bibfield  {author} {\bibinfo {author} {\bibfnamefont {M.}~\bibnamefont
  {Pant}}, \bibinfo {author} {\bibfnamefont {H.}~\bibnamefont {Krovi}},
  \bibinfo {author} {\bibfnamefont {D.}~\bibnamefont {Towsley}}, \bibinfo
  {author} {\bibfnamefont {L.}~\bibnamefont {Tassiulas}}, \bibinfo {author}
  {\bibfnamefont {L.}~\bibnamefont {Jiang}}, \bibinfo {author} {\bibfnamefont
  {P.}~\bibnamefont {Basu}}, \bibinfo {author} {\bibfnamefont {D.}~\bibnamefont
  {Englund}}, \ and\ \bibinfo {author} {\bibfnamefont {S.}~\bibnamefont
  {Guha}},\ }\href {\doibase 10.1038/s41534-019-0139-x} {\bibfield  {journal}
  {\bibinfo  {journal} {npj Quantum Information}\ }\textbf {\bibinfo {volume}
  {5}},\ \bibinfo {pages} {25} (\bibinfo {year} {2019})}\BibitemShut {NoStop}%
\bibitem [{\citenamefont {Chakraborty}\ \emph {et~al.}(2019)\citenamefont
  {Chakraborty}, \citenamefont {Rozpedek}, \citenamefont {Dahlberg},\ and\
  \citenamefont {Wehner}}]{chakraborty_routing}%
  \BibitemOpen
  \bibfield  {author} {\bibinfo {author} {\bibfnamefont {K.}~\bibnamefont
  {Chakraborty}}, \bibinfo {author} {\bibfnamefont {F.}~\bibnamefont
  {Rozpedek}}, \bibinfo {author} {\bibfnamefont {A.}~\bibnamefont {Dahlberg}},
  \ and\ \bibinfo {author} {\bibfnamefont {S.}~\bibnamefont {Wehner}},\ }\href
  {https://arxiv.org/abs/1907.11630} {\enquote {\bibinfo {title} {Distributed
  routing in a quantum internet},}\ } (\bibinfo {year} {2019}),\ \Eprint
  {http://arxiv.org/abs/1907.11630} {arXiv:1907.11630 [quant-ph]} \BibitemShut
  {NoStop}%
\bibitem [{\citenamefont {Mezzadri}(2007)}]{haarmatrices_mezzadri}%
  \BibitemOpen
  \bibfield  {author} {\bibinfo {author} {\bibfnamefont {F.}~\bibnamefont
  {Mezzadri}},\ }\href
  {https://www.ams.org/notices/200705/fea-mezzadri-web.pdf} {\bibfield
  {journal} {\bibinfo  {journal} {Notices of the American Mathematical
  Society}\ }\textbf {\bibinfo {volume} {54}},\ \bibinfo {pages} {592}
  (\bibinfo {year} {2007})}\BibitemShut {NoStop}%
\end{thebibliography}%
\end{document}